\documentclass[letterpaper, 10pt, conference]{IEEEtran}

\IEEEoverridecommandlockouts

\usepackage{amsmath}
\usepackage{url}
\usepackage{enumitem}
\usepackage{listings}

\usepackage{graphicx}

\DeclareGraphicsExtensions{.pdf,.jpeg,.png}
\graphicspath{{png/}}
\DeclareGraphicsExtensions{.pdf,.jpeg,.png}

\usepackage[caption=false,font=normalsize,labelfont=sf,textfont=sf]{subfig}
\usepackage[hidelinks]{hyperref}

\usepackage{url}
\usepackage{balance}

\usepackage{listings}
\usepackage[noend]{algpseudocode}
\usepackage{algorithm}
\usepackage{tabulary}

\usepackage{array}

\newcommand{\head}[2]{\multicolumn{1}{>{\centering\arraybackslash}p{#1}}{\textbf{#2}}}
\usepackage{color, colortbl}
\definecolor{LightCyan}{rgb}{0.88,1,1}
\definecolor{Gray}{gray}{0.9}

\usepackage{booktabs}   

\usepackage{draftwatermark}
\SetWatermarkScale{1}
\SetWatermarkColor[gray]{0.9}

\begin{document}


\title{FECBench: A Holistic Interference-aware Approach for Application Performance Modeling}
\author{\IEEEauthorblockN{Yogesh D. Barve\IEEEauthorrefmark{1}, Shashank Shekhar\IEEEauthorrefmark{2}\IEEEauthorrefmark{3}\thanks{\IEEEauthorrefmark{3}Work done as part of doctoral studies at Vanderbilt University}, Ajay Chhokra\IEEEauthorrefmark{1}, Shweta Khare\IEEEauthorrefmark{1},\\ Anirban Bhattacharjee\IEEEauthorrefmark{1}, Zhuangwei Kang\IEEEauthorrefmark{1}, Hongyang Sun\IEEEauthorrefmark{1}
and Aniruddha Gokhale\IEEEauthorrefmark{1}}

\IEEEauthorblockA{
	\IEEEauthorrefmark{1}Vanderbilt University, Nashville, TN 37212, USA and
		\IEEEauthorrefmark{2}Siemens Corporate Technology, Princeton, NJ, 08540, USA}
\IEEEauthorblockA{Email: \IEEEauthorrefmark{1}\{yogesh.d.barve,ajay.d.chhokra,shweta.p.khare,anirban.bhattacharjee,\\zhuangwei.kang,hongyang.sun,a.gokhale\}@vanderbilt.edu
and \IEEEauthorrefmark{2}shashankshekhar@siemens.com}
}

\maketitle


\begin{abstract}
Services hosted in multi-tenant cloud platforms often encounter
performance interference due to contention for non-partitionable resources,
which in turn causes unpredictable behavior and degradation in  application
performance.    To grapple with these problems and to define effective resource
management solutions for their services,  providers often must expend
significant efforts and incur prohibitive costs in developing performance
models of their services under a variety of interference scenarios on different
hardware. This is a hard problem due to the wide range of possible
co-located services and their workloads, and the growing heterogeneity in the
runtime platforms including the use of fog and edge-based resources, not to
mention the accidental complexity in performing application profiling
under a variety of scenarios.  To address these challenges,  we present
FECBench (Fog/Edge/Cloud Benchmarking), an open source framework
comprising a set of 106 applications covering a wide range of application
classes to guide providers in building performance interference prediction
models for their services without incurring undue costs and efforts.
Through the design of FECBench, we make the following contributions.
First, we develop a technique to build resource
stressors that can stress multiple system resources all at once in a controlled
manner, which helps to gain insights into  the impact of interference on an
application's performance.   Second, to overcome the need for exhaustive
application profiling, FECBench intelligently uses the design of experiments
(DoE) approach to enable users to build surrogate performance models of
their services.   Third, FECBench maintains an extensible knowledge base of
application combinations that create resource stresses across the multi-dimensional
resource design space. Empirical results using real-world scenarios to validate the
efficacy of FECBench show that the predicted application performance
has a median error of only 7.6\% across all
test cases, with 5.4\% in the best case and 13.5\% in the worst case.

\end{abstract}

\begin{IEEEkeywords}
Multi-tenant clouds, Performance Interference, Resource Management, Benchmarking.
\end{IEEEkeywords}

\section{Introduction}
\label{sec:intro}

\paragraph*{\textbf{Context}}
Multi-tenancy has become the hallmark of public cloud computing systems, where
physical resources such as CPU, storage and networks are virtualized and
shared among multiple different and co-located applications (i.e., tenants) to
better utilize the physical resources.
Although virtualization technologies such as virtual machines and  containers
allow cloud providers to increase the degree of multi-tenancy while still
providing isolation of resources among the tenants, there exist
non-partitionable physical resources such as the caches, TLBs, disk, and
network I/O, which are susceptible to resource contention thereby causing
adverse performance interference effects on the co-located
tenants~\cite{xavier2016understanding,iosup2011performance}.
Consequently, effective resource  management
solutions are required that can limit the impact of performance  interference
to acceptable levels such that the  service level objectives (SLOs) of the
applications can be
maintained~\cite{delimitrou2013paragon,berekmeri2016feedback,rattihalli2018exploring,delvalle2016exploring}.

\paragraph*{\textbf{Challenges}}
Developing effective resource management solutions (e.g., schedulers) requires
an accurate understanding of the target application's performance under
different application co-location scenarios so that the impact of performance
interference can be calibrated and accounted for in the 
solutions.   Recent studies~\cite{xu2018pythia, novakovic2013deepdive} have
built performance interference profiles for applications using a variety of
resource utilization metrics. Since performance interference is caused due to
the sharing of one or more non-partitionable resources, performance models
for the application-under-study (i.e., the target application) that account for
interference are developed by co-locating them with a variety of
\textit{resource stressor applications} (i.e., those applications that put
varying levels of pressure on the non-partionable resources) and recording the
delivered performance to the target application.

Creating such performance models, however, requires the developer to expend
significant efforts into application benchmarking and analyze application
performance under varying levels of resource stress.  Since the overall system
utilization is a function of the stresses imposed on multiple
types of resources in the system and the presence of multiple resources
represents a multi-dimensional space, creating varying levels of resource
stresses spanning this large design space is a difficult task.  Effective resource
management solutions, however,  require application performance models that
incorporate the impact of stresses on multiple resources all at once.
Although existing resource stressors,  such as \textit{dummyload} or
\textit{stress-ng} \cite{urlstressng}, provide  users with the control knobs to
exert the desired level of stress on a  resource, such as CPU or memory, these
tools operate on only one resource at a time.  Unfortunately, it is hard for
users to define the right kinds of application workloads that will create the
right levels of resource stress across the multi-dimensional space.
All these problems are further exacerbated with the addition of fog and edge
computing resources, which illustrate both increased heterogeneity and
constraints on resources, and where the performance interference effects may be
even more pronounced
\cite{brogi2017qos,jonathan2017nebula}.

Although, some frameworks/benchmarks  exist that can assist in the building of
the performance models,  these tools remain mostly disparate and it takes a
monumental effort on the part of the  user to bring these disparate tools
together into a single framework~\cite{iosup2014iaas}. Even then, such a
combined framework  may not be easy to use.  Moreover, a general lack of any
systematic approach to conduct the performance modeling process will force the
user to rely on \emph{ad hoc} approaches, which hurts reproducibility and leads
to reinvention of efforts~\cite{oleksenko2017fex}, not to mention the
possiblity of the resulting models missing out on critical insights.

Beyond these challenges, one question still persists: \emph{When is a
  performance model considered good enough such that it will enable effective
  resource management solutions?} In other words, how much application
profiling is required to build these performance models? One strawman strategy
to profile the application is to subject it to all possible resource
stresses. However, such an approach will be time-consuming and even infeasible
given the large number of combinations that can be executed on the different
resource dimensions, the variety in the co-located application types, and their
different possible workloads. Hence, there is a need for an intelligent
application profiling strategy that minimizes the profiling effort and thereby
the time and cost, while still providing sufficient coverage across all the
resources that contribute to application performance interference.
Unfortunately, there is a general lack of benchmarks and frameworks that
can aid the user in developing these models.

\paragraph*{\textbf{Solution Approach}}
To address these challenges,  we present FECBench (Fog/Edge/Cloud
Benchmarking), an open source framework comprising a set of 106
applications that cover a wide range of application classes to guide providers
in building performance models for their services without incurring undue costs
and efforts. The framework can then be used to predict interference levels and make
effective resource management decisions.  Specifically, through the design of FECBench,
we make the following contributions:
\begin{enumerate}
\item FECBench builds resource stressors that can stress
multiple system resources all at once in a controlled manner.  These resource
stressors help in understanding the impact of interference effects on an
application's performance.
\item To overcome the need for exhaustive
application profiling, FECBench intelligently uses the design of experiments
(DoE) approach to enable developers to build surrogate performance models of
their services.
\item FECBench maintains an extensible knowledge base of
application combinations that create resource stresses across the multi-dimensional
resources design space.
\end{enumerate}

Empirical results using real-world scenarios for validating the efficacy of FECBench
show that the predicted application performance 
has a median error of only 7.6\% across all test cases, with 5.4\% in
the best case and 13.5\% in the worst case. A short poster version
of this paper describing initial works can be found in \cite{barve2018fecbench}.

\paragraph*{\textbf{Paper Organization}}
The rest of the paper is organized as follows:
Section~\ref{sec:motiv} delves into the details of performance interference,
surveys the literature in this realm; Section~\ref{sec:goals_challenges}
elicits the key requirements for a solution such as FECBench;
Section~\ref{sec:methodology} presents the design and implementation of
FECBench, explaining how it meets the requirements outlined earlier;
Section~\ref{sec:validate} presents an extensive set of results
validating the different features of FECBench;
and finally Section~\ref{sec:conclusions} presents concluding remarks
discussing the implications of using FECBench and alluding to future
work.

\section{Background and Literature Survey}
\label{sec:motiv}
In this section, we provide details on performance interference and its impact on
application performance.  We then present a survey of the literature in this
area and the limitations of existing approaches, which motivates the key requirements of our FECBench
solution.

\subsection{Sources of Interference and Impact on Performance}
\label{sec:soi}

\begin{figure}[htb]
\centering
\includegraphics[width=1.0\linewidth,height=4cm]{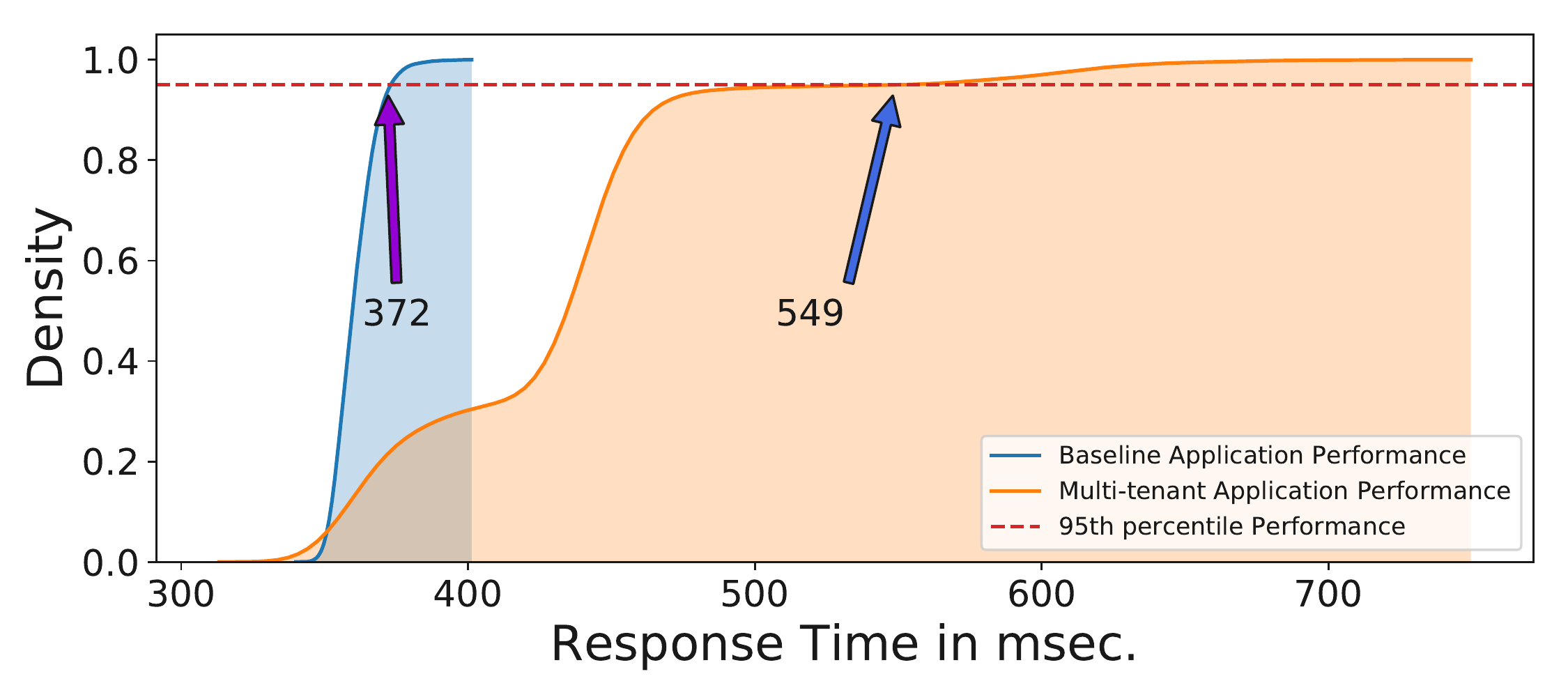}
\vspace*{-9mm}
\caption{CDF representation of prediction inference response times for the
  Inception RESNETv2 Keras model. We can see that the application's performance degrades when co-located with background
  applications as compared to its performance when running in isolation.}
\label{fig:resnet}
\end{figure}

Co-located applications on the same physical resources of a cloud platform will impose
varying degrees of pressure (stress) on the underlying resources.  When these
resources are hard to partition or are isolate, the contention for these
resources will cause interference effects on the executing applications and
degrade their performance.  For compute-intensive applications, resources such
as CPU core and Last Level Cache (LLC) can cause
interference. Similarly, for communication-intensive applications, resources such as memory bandwidth, disk I/O, and network
can cause interference. For example, Figure~\ref{fig:resnet}
shows the performance degradation of an application that uses
the Inception RESNETv2 deep learning model \cite{urlKeras}.  The figure illustrates a
cumulative distribution function (CDF) for the 95th percentile response times of the application as its SLO with and
without interference. Due to the significant difference in the observed response times, it is important for
resource management solutions to incorporate the impact of interference to
maintain application SLOs.

\subsection{Related Work}
\label{sec:relwork}
We now present prior efforts that focus on quantifying
and modeling performance interference and classify these along three
dimensions.

\paragraph*{\textbf{Interference Quantification}}
\label{sec:soi_quantification}

Two fundamentally different approaches to quantifying performance interference
have been reported in the literature.  The \emph{Bubble-Up}
approach~\cite{mars2011bubble} measures \emph{sensitivity} (i.e., impact of
co-located applications on the target application) and \emph{pressure} (i.e., impact
of the target application on co-located applications) using a
synthetic stressor application called \textit{bubble}. The bubble generates a
tunable amount of pressure on a given resource, such as memory or LLC. With the pressure applied on a resource, the target application is
executed simultaneously with co-located applications, and its performance
metrics such as the completion time are measured. This experiment is repeated
for different pressure levels in both the memory and the cache
subsystems.  Although this approach is effective, the bubble is limited to
memory sub-system only and the approach is limited to two co-located
applications.
Yang et. al. \cite{yang2013bubble} extended Bubble-Up to allow performance
interference beyond two co-located applications and other shared resources such
as network, I/O, cores, etc.  The observed application performance
degradation is used to construct sensitivity and pressure
profiles which are used to determine if a given co-location will cause
degradation in the performance of an application.

A different approach is presented in \emph{DeepDive}~\cite{novakovic2013deepdive}, in
which the performance interference is predicted based on the aggregate
resource system utilization on the running system. Unlike Bubble-Up, where
performance of an application is measured for stress levels independently on
each resource, in DeepDive, application interference is measured by monitoring
resource usage statistics of all co-located applications. In DeepDive, an
application running inside a virtual machine is placed on an isolated physical
machine and the resource utilization
statistics  are measured. The application is then migrated/placed on a physical
machine by estimating the quality of interference level on the application and
the co-located running applications.
%
To model the application performance, we employ a technique used in
DeepDive that utilizes the system resource metrics to build performance
models.

\paragraph*{\textbf{Interference-aware Predictive Modeling}}
\label{sec:rel_interference_analysis}

Building performance interference models and using them to predict the expected
levels of interference for a given co-location configuration and workloads is
important.  Paragon \cite{delimitrou2013paragon} presents an interference-aware
job scheduler, in which an application's performance is predicted using
collaborative filtering.  The performance prediction model is built using
performance data that is measured by subjecting the test application against
individual resource stressors that stress only one resource at a time.
In comparison, FECBench takes into account the cumulative effect of all the
resources to  build an interference prediction model.

Zhao et. al.~\cite{zhao2013empirical} studied the impact of co-located
application performance for a single multi-core machine. They developed a
piecewise regression model based on  cache contention and bandwidth
consumption of co-located applications. Similarly, their work captures the
aggregate resource utilization of two subsystems, namely, cache  contention and
memory bandwidth, in determining the performance degradation. Our approach also
considers disk and CPU resources in building prediction models and is not
restricted to any hardware.

In DIAL \cite{javadi2017dial}, interference detection is accomplished using
decision tree-based classifier to find the dominant source of resource
contention. To quantify the resource interference impact on a webserver
application's  tail response, a queuing model is utilized to determine the
application's response time under contention. To minimize the effects of
interference, it proposed using a runtime controller responsible for dynamic
load-balancing of queries from webserver.  Subramanian et.
al.~\cite{subramanian2015application} presented an application slowdown model, 
which estimates the application performance with high accuracy using cache
access rate and memory bandwidth. However, the system was validated using a
simulator and not on real hardware system. In contrast, FECBench is
geared towards real hardware.

The ESP project~\cite{mishra2017esp} uses a two-stage process for interference
prediction. It first performs feature extraction, and then builds regression
model to  predict performance interference. It creates separate models for each
co-location groups. Also, its training data workload consists of all the
possible applications that can run in the cluster. It then collects performance
data for some combinations out of all the possible combinations to build the
interference model. Similarly, Pythia \cite{xu2018pythia} describes an approach
for predicting resource contention given a set of co-located workloads. Both
ESP and Pythia assume that they have \emph{a priori} information of all
possible running workloads, based on which an interference model is created for
a new application. In comparison, FECBench relies on the performance metrics
obtained when co-located with a fixed number of resource stressors and does not
need to have prior information of all the running applications in the cluster.

\paragraph*{\textbf{Interference-related Synthetic Benchmarks}}
One of the major roadblocks when investigating and building the performance
interference modeling is the lack of representative benchmarking applications.
Cuanta~\cite{govindan2011cuanta} built a synthetic cache loader that emulates
pressure for varying tunable intensities on the LLC resource. It supports a
Linux kernel module that invokes hypervisor system call to create the desired
level of memory utilization. This kernel module resides inside a virtual
machine.  In contrast, our approach is non-intrusive and does not require any changes
to the Linux kernel.

iBench~\cite{delimitrou2013ibench} developed an extensive set of synthetic
workloads that induce pressure on different resource subsystems. These
workloads are built in a way to exert tunable utilization pressure on system
resources such as L1, L2, iTLB, memory, LLC, disk, network, etc. independently.
In~\cite{pu2013your}, synthetic workloads were used to create pressure
on network and CPU systems. Bubble-up~\cite{mars2011bubble}, which was
described earlier, is another effort in this category.

While these efforts made a step in the right direction, most existing approaches 
rely on manual tuning of the resource stressors to create the desired
level of stress. As a result, prior approaches cannot adapt to changes in the
underlying architecture. In contrast to earlier works, our approach finds
application pairs and creates a resource stressor knowledge-base in an
automated fashion. As a result, our approach can adapt to changes in the
underlying architecture and can be reused. Moreover, with the help of design of
experiments, FECBench reduces the profiling effort of the applications.

\section{Solution Requirements and Proposed Approach}
\label{sec:goals_challenges}

Based on the literature survey and unresolved challenges, we derive the
following requirements for FECBench.

\textbf{1. Benchmarking with Ease}: Benchmarking and profiling applications can
be a very tedious task because it involves configuration of probes on resources
such as CPU, network or disk, and collection of many hardware- and
application-specific performance metrics \cite{iosup2014iaas,silva2013cloudbench,jayathilaka2018detecting}.
In this regard, tools such as \texttt{CollectD} \cite{urlCollectd} and \texttt{Systat}
allow monitoring and collecting system metrics. Often, more than one tool may
be required to collect the metrics of interest, which makes it hard for the
user to integrate the tools. Moreover, dissemination of the monitored metrics
in a timely manner to a centralized or distributed set of analysis engines must
be supported to build performance interference models of the applications.

To address these challenges and to make the task of benchmarking easier and
intuitive for the user, FECBench uses higher-level, intuitive abstractions in
the form of domain-specific modeling \cite{anirbanUCC,barve2018pads} and generative techniques to synthesize
the generation of configurations, metrics collection and dissemination.
Our recent work~\cite{barve2018upsara} describes these capabilities and hence it is not
a focus of this paper but we discuss this requirement for completeness sake.

\textbf{2. Automated Construction of Resource Stressors}:
Tools like lookbusy and stress-ng can be utilized to create resource stress on
CPU in a controlled tunable manner. Similarly, tools like iPerf can be utilized
to create resource stress on the network resource. Despite this, there is a lack
of open-source tools that can stress multiple resources simultaneously, also in a tunable manner.  Moreover, some of the resource stressors are
platform-specific, which hinders their applicability to heterogeneous
platforms. Prior studies have presented design of stressors, which requires a deep  understanding of the
underlying hardware architecture  and low-level resource
characteristics. Acquiring the skills to utilize these tools thus incurs a
steep learning curve.

To address these concerns, Section~\ref{sec:methodology} presents a process
pipeline with offline and online stages that construct the multi-resource
stressors in an automated fashion by leveraging machine learning techniques.

\textbf{3. Minimizing the Prohibitive Profiling Cost}:
When building the performance interference model, a user must profile the
application's performance metrics against different configurations of resource
utilization on the running system. However, since we have multiple resources,
the resource utilization can be seen as a multi-dimensional design space.
One approach to profiling is to cover exhaustively the entire design space and
obtain the performance metrics for the application.  However, the cost and
time for executing these experiments will be very high.  Thus, there
is a need to significantly reduce the profiling effort while deriving good
performance interference models.

To that end, we use the \emph{design of experiments} (DoE) approach in Section \ref{sec:doeLatin}
to explore the multi-dimensional resource metrics for building the performance interference
models.  DoE is a statistical technique which has been used  to substantially
lower the number of experimental runs required to collect data.
In our study, we leverage the Latin Hypercube Sampling (LHS) DoE technique.

\section{FECBench Methodology}
\label{sec:methodology}

We now present FECBench and demonstrate its design methodology by building a
performance interference model and explaining each step of its design.

\subsection{FECBench Methodology and its Rationale}
\label{fecbench_pipeline}

\begin{figure}[htb]
\centering
\includegraphics[width=1.0\linewidth,height=5.5cm]{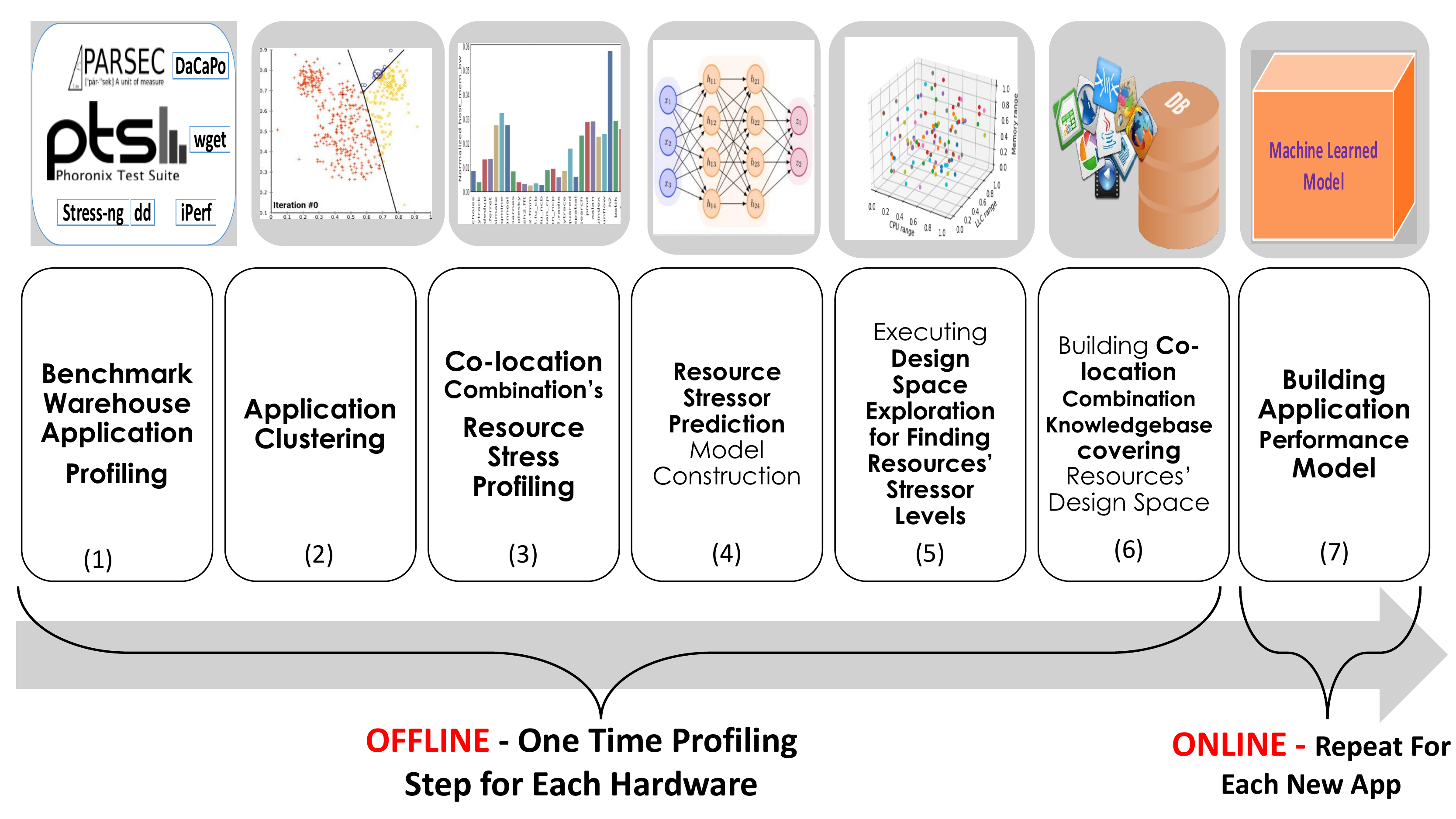}
\vspace*{-5mm}
\caption{FECBench methodology.}
\label{fig:pipeline}
\end{figure}

Figure~\ref{fig:pipeline} presents the FECBench process.  The rationale for
this process is described below and details of each step follow.

Recall that the goal of FECBench is to minimize the efforts for developers in
building interference-aware performance models for their applications by
providing them a reusable and extensible knowledge base. To that end, FECBench
comprises an offline stage with a set of steps to create a knowledge base
followed by an online stage. Developers can use the same offline stage process
to further refine this knowledge base.

\begin{figure*}[htb]
\centering
\includegraphics[width=1.0\linewidth,height=4cm]{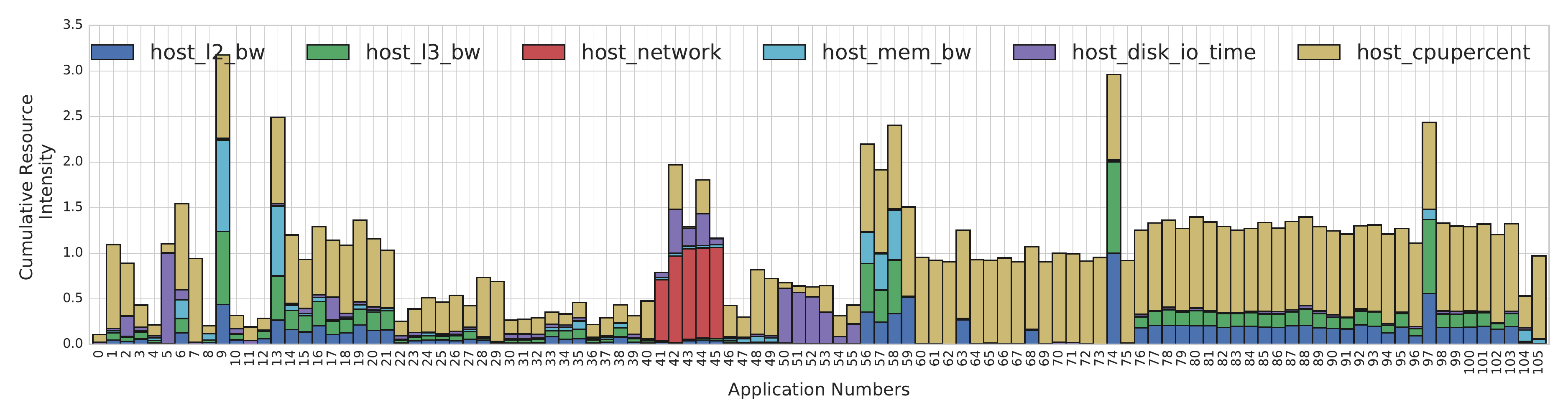}
\vspace*{-9mm}
\caption{Utilizations of different resources for each of the 106 applications from the benchmarking warehouse.
Each application is represented by a unique number and is depicted along the horizontal axis.}
\label{fig:resourceUtil}
\end{figure*}

Accordingly, the first step (\ref{appsdetails}) of the offline stage defines a \textit{Benchmark
  Warehouse} (BMW), which is a collection of resource utilization metrics
obtained by executing a large number and variety of applications  on a specific
hardware and measuring the impact on each resource type independently.  The
second step (\ref{sec:clustering}) clusters these applications according to their similarity in how
they stress individual resources.  Clustering minimizes the unwieldiness
stemming from the presence of a large number of application types in the
performance model building process.   Since we are interested in performance
interference, the knowledge base must capture the stress on resources stemming
from executing a combination of co-location patterns of applications belonging
to the different clusters found in the earlier step (Step 3 \ref{sec:colocutil}).

Using all this data, we define a resource stressor prediction model (Step 4 \ref{sec:utilmodel}),
which can be used to predict the expected stress along the multi-dimensional
resource space given a new co-location pattern. Since such a model building
process itself may need exhaustive searching through every possible combination
of resource stresses along the multi-dimensional resource space, we create
surrogate models using design of experiments (DoE), specifically, the Latin Hypercube
Sampling (LHS) approach (Step 5 \ref{sec:doeLatin}).

The reduced search strategies of Step 5 give rise to a knowledge base (Step 6 \ref{knowledgebase}),
which is then used in the online stage (Step 7 \ref{interfmodel}) that stresses a target
application across different resource utilization regions from the design space
to train a model for that target application and utilize it to predict its
performance at runtime. The developer of a new application need only
conduct Step 7 while leveraging all the previous steps.  If a completely new
hardware configuration is presented, the knowledge base must be updated by
repeating all the steps of the offline stage. The steps are detailed next.

\subsection{Benchmarking Isolated Characteristics of Applications}
\label{appsdetails}

To build stressors that can stress different resources of a system, we first
study the characteristics of different applications in isolation.  We have
profiled 106 applications from existing but disparate benchmarking suites, such
as PARSEC~\cite{bienia2008parsec}, DaCaPo~\cite{blackburn2006dacapo},
PHORONIX~\cite{urlPhoronix}, STRESS-NG~\cite{urlstressng}, as well as networked
file-server applications, which collectively form our \textit{benchmarking
  warehouse} (BMW).  These applications represent a diverse range spanning from
cloud computing to approximate computing
workloads~\cite{yazdanbakhsh2017axbench}.

In this step, we profile the applications by running each in isolation on a
system so as to document the resource utilization imposed by that
application.  Let $A$ denote the set of all applications in BMW. For each
application $a\in A$, we collect its runtime utilization metrics on the host
system when run in isolation, including CPU, L2/L3 cache bandwidth, memory
bandwidth, disk, network, etc. For a total number $R$ of resources considered,
the vector $U(a) = [u^{(1)}(a), u^{(2)}(a), \dots, u^{(R)}(a)]$ is then logged
in a database, where $u^{(r)}(a)$ denotes the utilization on a particular
resource $r \in \{1, 2, \dots, R\}$ when running the application.
Figure~\ref{fig:resourceUtil} presents the resource utilization characteristics
of 106 applications. The experiment host used is Intel(R) Xeon(R)
CPU E5-2620 v4 machine with 16 physical cores. As we can see from the figure,
the applications exhibit a high degree of coverage across the resource
utilization spectrum for the different system resources.
\subsection{Application Clustering}
\label{sec:clustering}
Given the large number of applications available in the BMW, it is likely that
some of them exhibit similar characteristics with respect to the
resource utilizations. 
For example, applications with numbers 80 and 81 in Figure~\ref{fig:resourceUtil} have similar utilizations
with respect to CPU, L2 bandwidth and L3 bandwidth.
This step performs clustering to identify those applications that share similar
resource utilization characteristics.  Moreover, application clustering allows
us to select only a subset of applications from the BMW for the subsequent
co-location resource utilization study. This helps to significantly reduce the
number of application combinations that need to be profiled and tested.

\begin{figure}[htb]
\centering
\includegraphics[width=1.0\linewidth,height=10cm]{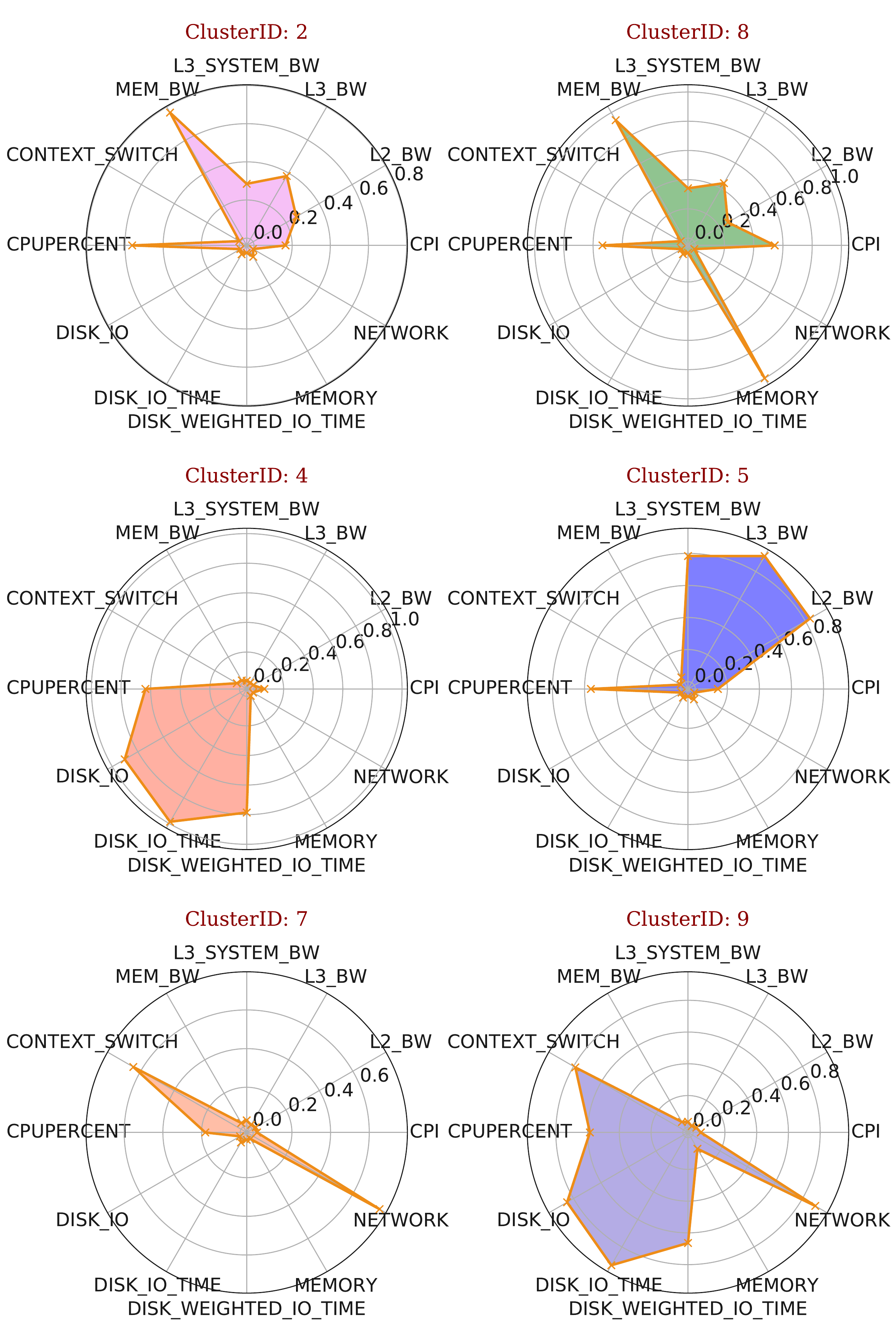}
\vspace*{-5mm}
\caption{Radar charts illustrating the resource profiles for
clusters 2, 4, 5, 7, 8 and 9. Resource pressure is higher when the vertices are
closer to the edge of the radar chart on the resource axis.}
\label{fig:appcluster}
\end{figure}

Machine learning approaches, such as $K$-means clustering or Support Vector
Method-based clustering, have been commonly used to find similarities in
datasets~\cite{mishra2010towards}.  In this study, we leverage the $K$-means
algorithm to cluster all applications from the BWM in the $R$-dimensional
space, where each dimension represents the utilization from a particular
resource $r\in \{1, 2, \dots, R\}$. Thus, each application $a \in A$ is represented by a
point $U(a) = [u^{(1)}(a), u^{(2)}(a), \dots, u^{(R)}(a)]$ in the
$R$-dimensional space.  We use the Silhouette algorithm
\cite{rousseeuw1987silhouettes} to determine the ideal number of clusters. For
the considered 106 applications, running the algorithm leads to $K = 13$
clusters. Figure~\ref{fig:appcluster} shows the resource utilization characteristics for some of these clusters. As can be seen, Cluster 5 is
L3, L2 and L3 system bandwidth-intensive. Similarly, Cluster 2 is memory bandwidth
intensive. Cluster 9 shows high utilization pressures across network, disk and CPU.

\subsection{Resource Utilization Profiling for Co-located Workloads}
\label{sec:colocutil}

\begin{figure}[thb]
\centering
\includegraphics[width=1.0\linewidth,height=6cm]{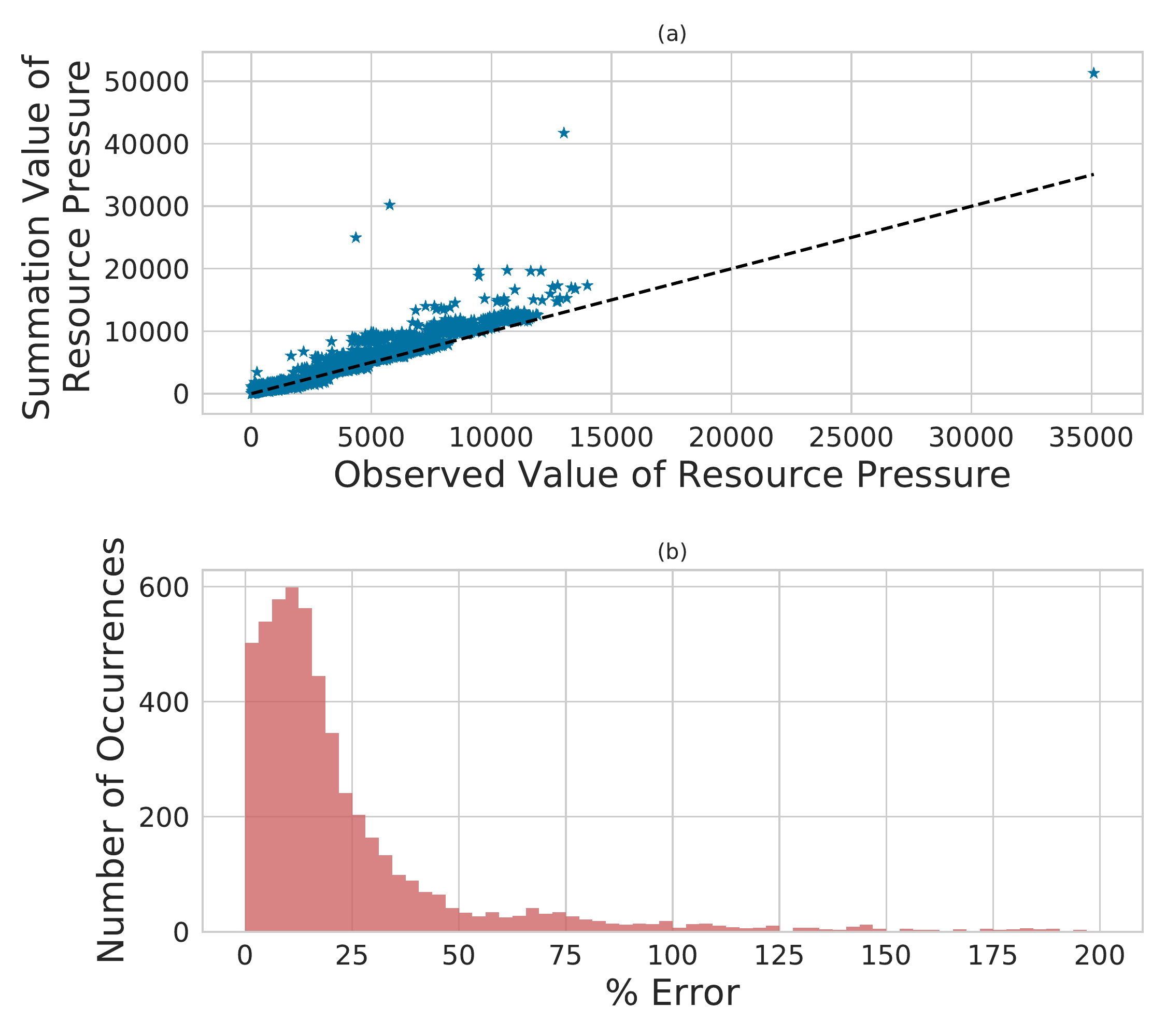}
\vspace*{-5mm}
\caption{Chart (a) shows the summation of the isolated L3 bandwidth resource
  pressures and the observed resource pressure on the system. Chart (b) shows
  the distribution of error from direct summation of isolated resource
  pressures of the applications. The mean absolute percentage error is
  47\%.}
\label{fig:sum_host_L3_BW}
\end{figure}

Since running a single application may not create the desired stress levels for
the system resources, we are interested in finding those application mixes that
together can create a more diverse set of resource stress levels.  We first
observe from our empirical experiments that the utilization of a resource on a
system by running a set of co-located applications cannot be obtained by simply
summing the resource utilizations of these applications when executed in
isolation. This is validated by Figure~\ref{fig:sum_host_L3_BW},
which shows that a direct summation of the isolated applications' L3 bandwidth
utilizations incurs significant difference margins, with a mean absolute percent error of 47\%.
Similar behavior is also seen for other system resources.

This calls for profiling the resource utilization characteristics of different
application co-location patterns.  However, empirically running all application
combinations is extremely time consuming.  This step and the next together
build a resource prediction model that determines the resource utilizations
for any given application co-location pattern.

Before building a
resource prediction model, we collect resource utilization data in this step by
profiling a selection of application mixes, similar to the way we profiled a
single application in Section~\ref{appsdetails}.   Specifically, we pick an
arbitrary application (e.g., the centroid) from each of $K$ clusters and
co-locate applications from different clusters to create different resource
stressors.   Let $d^{\max}$ denote the maximum number of co-located
applications that are allowed in a run.  This gives a total of
$\sum_{d=1}^{d^{\max}} \binom{K}{d}$ application combinations.

Depending on the execution time needed to perform the profiling, one can choose
a random subset of these combinations for building the prediction model.  In
our experiment, we have $K = 13$ and $d^{\max} = 8$ (since we are using a
16-core server and each application executes on 2 cores), which gives a total
of 7,098 combinations. Among them, we profiled around 2,000 combinations for
building the prediction model.


\subsection{Building Resource Stressor Prediction Model}
\label{sec:utilmodel}

Based on the resource utilization data collected from the last step, this step builds a resource stressor prediction model.
We leverage random forest regression to determine the individual resource utilization given a set of co-located applications.
Random forest regression is a widely used non-parametric regression technique for capturing non-linearity in the dataset.
Unlike other methods, it performs well even with a large number of features and a relatively small training dataset, while providing
an assessment of variable importance. Random forest is an ensemble-learning-based approach, where the results of multiple models, in this case,
decision trees, are averaged to provide the final prediction.
A decision tree recursively partitions a sample into increasingly more homogeneous groups
up to a pre-defined depth. Terminal nodes in the tree then contain the final result.
Random forest randomizes the creation of decision trees in two
ways: 1) each decision tree is created from a random subset of input data; 2) each tree-partition is based on a random subset of input features.

For any application combination $A$, the following input features are used in the random forest prediction model:
\begin{itemize}
  \item A $K$-dimensional vector $C = [c_1, c_2, \dots, c_{K}]$, where each element $c_i$ takes the value 1 if an application from the $i$-th cluster is selected in the combination $A$ and 0 otherwise.
  \item A $R$-dimensional vector $U_{+}(A) = [u^{(1)}_{+}(A), u^{(2)}_{+}(A), \dots, u^{(R)}_{+}(A)]$, where each element $u^{(r)}_{+}(A)$ represents the sum of individual utilizations for the $r$-th resource from all applications in $A$, i.e, $u^{(r)}_{+}(A) = \sum_{a\in A} u^{(r)}(a)$.
\end{itemize}
The output is another $R$-dimensional vector that predicts the utilization for all resources when executing the application combination $A$, i.e., $\tilde{U}(A) = [\tilde{u}^{(1)}(A), \tilde{u}^{(2)}(A), \dots, \tilde{u}^{(R)}(A)]$.
Thus, the resource stressor prediction model maps the input to the output via a prediction function $f$ as follows:
\begin{align*}
\tilde{U}(A) \leftarrow f\big(C, U_{+}(A)\big)
\end{align*}

\subsection{Design of Experiments (DoE) Specification}
\label{sec:doeLatin}

To build a performance interference model for a target application when it is
co-located with other applications, we need to measure its performance on a
system that experiences different resource stress levels.  Due to the large
number of possible stress levels along multiple resource dimensions, it is not
practical to test all of them.  Therefore, we adopt the design of experiments (DoE)
\cite{cavazzuti2013design} approach by generating a small number of sample
points in the multi-dimensional space that maximizes the coverage of the
different resource utilizations.

To this end,
we leverage the Latin Hypercube Sampling (LHS) method
\cite{cavazzuti2013design} that generates sampled regions across the
$R$-dimensional resource utilization space.  Specifically, LHS divides each
resource dimension into \emph{M} equally-spaced intervals, and then selects $M$
sample intervals in the entire $R$-dimensional space that satisfies the Latin
Hypercube property: each selected sample is the only one in each axis-aligned
hyperplane that contains it. The LHS method has a clear advantage over random
sampling, which could potentially lead to selections of samples that are all
clumped into a specific region.  Moreover, the number $M$ of samples in LHS does
not grow with the number of dimensions. For a given choice of $M$, it generates
a collection $H = \{h_1, h_2, \dots, h_M\}$ of $M$ hypercubes. Since each
dimension represents the resource utilization of a corresponding resource in
our case, its overall range is $[0, 1]$. Therefore, each generated hypercube
$h_i\in H$ in a resource dimension $r$ has the range
$\left[\frac{x^{(r)}_i}{M}, \frac{x^{(r)}_i+1}{M}\right]$ for some $x^{(r)}_i
\in \{0, 1, \dots, M-1\}$. We refer interested readers to \cite{loh1996latin}
for an in-depth explanation of the LHS method. In our experiment, we set
\textbf{$M = 300$}.


\subsection{Creating the Stressor Knowledge Base}
\label{knowledgebase}

We now create a knowledge-base of applications and their workload mixes that map
to the different resource utilization levels as determined from the DoE
exploration.  Let $\mathcal{SA}$ denote the set of all application combinations generated
in Section \ref{sec:colocutil}.   We consider every application combination $A
\in \mathcal{SA}$  and use the resource stressor prediction model of Section \ref{sec:utilmodel} to predict its
utilization $\tilde{u}^{(r)}(A)$ for each individual resource $r\in\{1, 2,
\dots, R\}$.  We then fill up each of the $M$ hypercubes sampled in
Section~\ref{sec:doeLatin} with the application combinations that belong to it.
Specifically, for each application combination $A$, it is assigned to
hypercube $h_i\in H$, if $\tilde{u}^{(r)}(A) \in \left[\frac{x^{(r)}_i -
    \delta}{M}, \frac{x^{(r)}_i+1 + \delta}{M}\right]$ for all $1\le r\le R$,
where $\delta > 0$ is a tolerance parameter to extend the boundaries of the
hypercubes to account for the inaccuracy of the stressor prediction model. In our
experiment, we set $\delta = 0.1$.

\subsection{Building Performance Interference Prediction Model}
\label{interfmodel}


In the last step (which is an online step), a developer must construct a
performance interference prediction model for a new target application $b$ that
is introduced for the first time onto the platform.  The goal is to predict a
specified QoS metric $q$ for the target application when it is co-located with
any set $B$ of applications. To that end, we leverage a
regression-based Decision Tree model \cite{quinlan1986induction}. The input of the prediction model
is a $R$-dimensional vector $U = [u^{(1)}, u^{(2)}, \dots, u^{(R)}]$ showing the utilization of different system resources
before the target application $b$ is deployed. The output is the predicted QoS metric for the target application $b$, which we denote as
$\tilde{q}(b, U)$, under the current system utilization $U$.
Thus, the interference prediction model maps the input to the output via a prediction function $g$ as follows:
\begin{align*}
\tilde{q}(b, U) \leftarrow g\big(U\big)
\end{align*}

To build the regression model, the target application is executed under
different resource stress levels identified by the design of experiments in
Section~\ref{sec:doeLatin}.  For each of the $M$ resource stress levels, the
target application is executed along with a selected application combination
from the knowledge base that corresponds to the desired resource stress level.
To select the application combination, the closest application to the center of each hypercube is chosen.
This selected application combination is first run on the platform. After a warm-up period, the target application is then deployed on the same platform, and
its performance QoS metric is logged.  This process is repeated for all the $M$
resource stress levels.  The results are used as training data to train the
regression model above.
In FECBench, we consider the response time, which is the computation time of the
application, as the QoS metric used for latency-sensitive applications with soft
real-time requirements.

\section{System Architecture and Implementation}
\label{sec:implementation}

\begin{figure}[htb]
\centering
\includegraphics[width=\linewidth]{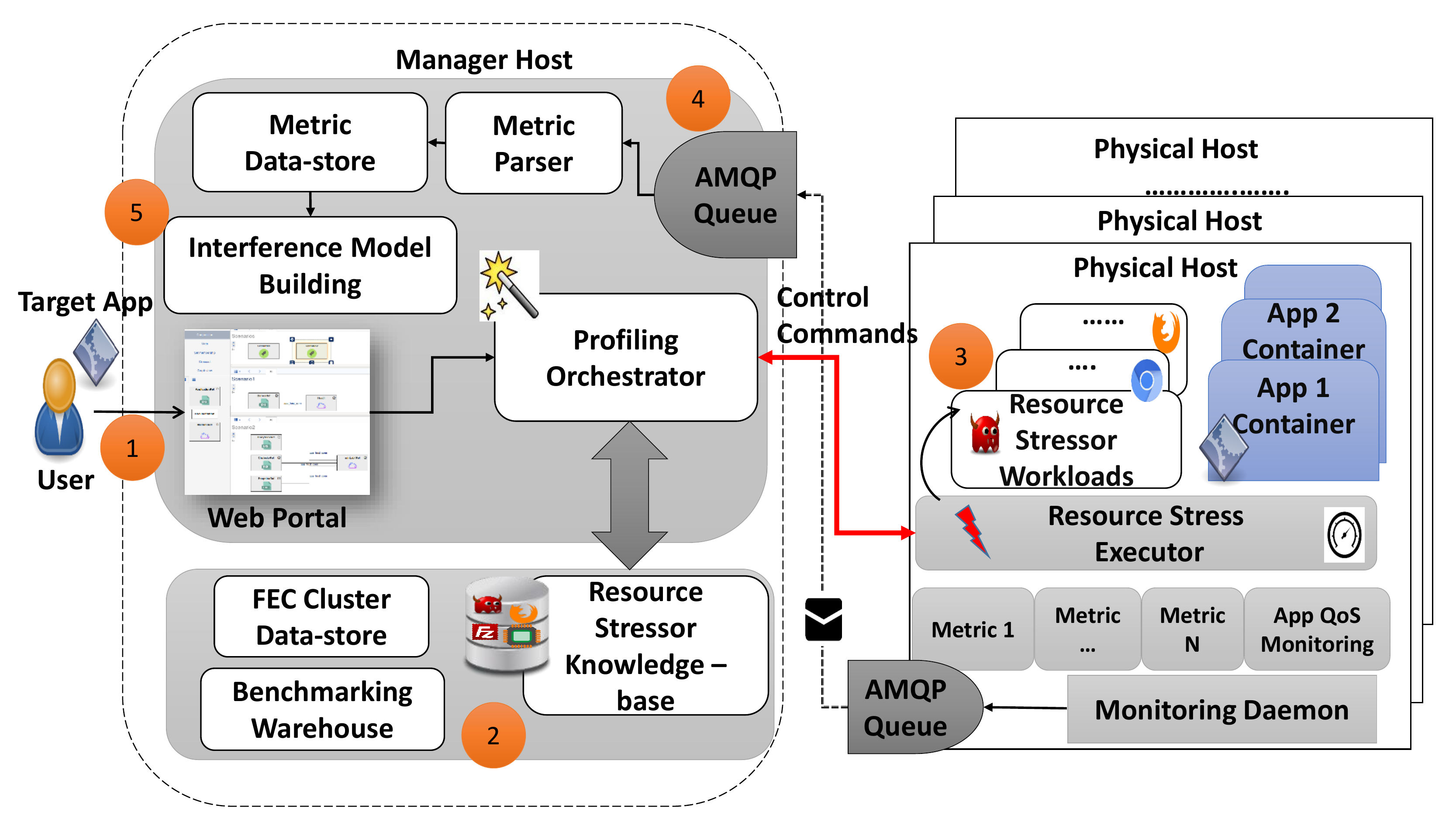}
\vspace*{-5mm}
\caption{FECBench in action. The figure shows different components of the FECBench along with the manager host and the target physical hosts on
which application profiling takes place.}
\label{fig:fecbench_action}
\end{figure}

Figure \ref{fig:fecbench_action} shows the different components of FECBench.
There are two classes of nodes: \emph{manager host} and
\emph{physical hosts}. Manager host is responsible for the management
and orchestration of FECBench.  We describe the numbered components in the
figure below.

In {\large \textcircled{\small 1}}, the \textit{Webportal} component at the
manager host allows the user to interact with FECBench. The Webportal is built
using a visual domain specific modeling language \cite{barve2016cloud}. It allows the user to submit
the target application whose performance interference model needs to  be
constructed. The Webportal initiates the profiling of the target application,
and relays this information to the \emph{Profiling orchestrator}, which  then fetches the required information -- resource stressors and system availability --
from the FECBench system information block represented by {\large \textcircled{\small 2}}.
Components in {\large \textcircled{\small 2}} comprise of the FEC Cluster datastore, Benchmarking Warehouse and
Resource Stressor Knowledgebase. The FEC Cluster datastore consists of the
current information about the cluster. Equipped with the required information,
the Profiling orchestrator deploys the target application on the desired
physical host, where the interference-aware profiling of the application
takes place. In {\large \textcircled{\small 3}}, the target application is
subjected to  various resource stressors as obtained from the stressor
knowledge base.  The monitoring probes on the physical hosts monitor the system
as well as the application QoS metrics and this information is relayed to the
manager host. In {\large \textcircled{\small 4}}, the desired metrics are
parsed and data is stored in a time series datastore. In {\large
  \textcircled{\small 5}}, FECBench constructs the interference-aware
performance model of the target application.

The monitoring of the system performance metrics is done using the
\emph{CollectD} monitoring program. For the system metrics not supported by CollectD, custom Python
based plugins are written that feed the metric data to the
CollectD daemon. CollectD plugins for the Linux \emph{perf} utility and
\emph{Likwid} monitoring tool \cite{urlLikwid} are written to monitor and log additional
system metrics such as the cache and memory level bandwidth
information. Docker-based resource stressor containers have also been
built. The monitored metrics are relayed in real time using the AMQP message
queues. Metric parsers for the gathered data is written using both Golang and
Python languages. We use \emph{Influxdb} to provide time series database for
storing the monitoring metrics \cite{urlInfluxdb}. For building performance models,  we 
leverage the machine learning libraries provided by the \emph{Scikit} library
in Python \cite{urlScikit}.

\section{Experimental Validation}
\label{evaluation}

This section validates the claims we made about FECBench.  To that end, we
demonstrate how FECBench enables savings in efforts in building the performance
models of applications and their accuracy in making resource management
decisions.  We validate the individual steps of the FECBench
process. We also present a concrete use case that leverages FECBench for
interference-aware load balancing of topics on a
publish-process-subscribe system.

\subsection{Experimental Setup}
\label{setup}
We validated the FECBench claims for a specific hardware comprising an Intel(R)
Xeon(R) CPU E5-2620 v4 compute node with 2.10 GHz CPU speed, 16 physical cores,
and 32 GB memory.  The software details are as follows: Ubuntu 16.04.3 64-bit,
Collectd (v5.8.0.357.gd77088d), Linux Perf (v4.10.17) and Likwid Perf
(v4.3.0).  For the experiments, we configured the \texttt{scaling\_governor}
parameter of CPU frequency to \texttt{performance} mode to achieve the maximum performance.

\subsection{Validating the Resource Stressor Prediction Model}
\label{eval_stressormodel}
To build and validate the resource stressor prediction model (Step 4 of
FECBench), we first need to obtain a dataset that includes the resource utilizations
under different application co-location scenarios (Step 3).
In our experiment, we pin each application to 2 cores of the test node for a
maximum of 8 co-located applications (since the node has 16 cores).  Our
offline profiling for Step 3 produced a dataset of about 2,000 data points, of
which we used 80\% for training, 10\% for testing and the remaining 10\%
for validation. Table~\ref{tab:stressor_table} illustrates the
performance of the resource stressor prediction model.  We see that the learned
models have high accuracy for both the test and the trained dataset. We used the
same accuracy measure, \textit{coeffecient of determination ($R^{2}$)},
as in prior studies~\cite{mishra2017esp}. We observe an accuracy of 99.1\% and 99.3\% for the training and testing
data, respectively, for memory bandwidth. The learned model also has low bias and variance since both
testing and validation errors converge for most cases.

\begin{table}[htb]
\vspace*{-2mm}
\caption{Performance of Learned Models for Resource Stressors}
\label{tab:stressor_table}
\centering
\scalebox{0.9}{
\begin{tabular}{lcccc}
\toprule
      \textbf{Feature} &  \head{1.2cm}{Test Accuracy } &  \head{1.2cm}{Train Accuracy } & \head{1.2cm}{Validation Accuracy}  \\
\midrule
  MEM\_BW &         99.305 &          99.073 &        98.930 \\
       \rowcolor{LightCyan}
   CPUPERCENT &         89.896 &          88.776 &        88.889 \\
       MEMORY &         98.310 &          98.346 &        86.143 \\
      \rowcolor{LightCyan}
        L3\_BW &         99.085 &          98.839 &        98.037 \\
      NETWORK &         99.663 &          99.665 &        99.673 \\
       \rowcolor{LightCyan}
 L3\_SYSTEM\_BW &         99.367 &          98.888 &        98.452 \\
        L2\_BW &         99.454 &          99.130 &        97.916 \\
       \rowcolor{LightCyan}
 DISK\_IO\_TIME &         88.002 &          88.464 &        88.519 \\
%
%
%
%
%
%
\bottomrule
\end{tabular}
}

\end{table}

\begin{figure}[htb]
\centering
\includegraphics[width=1.0\linewidth,height=6cm]{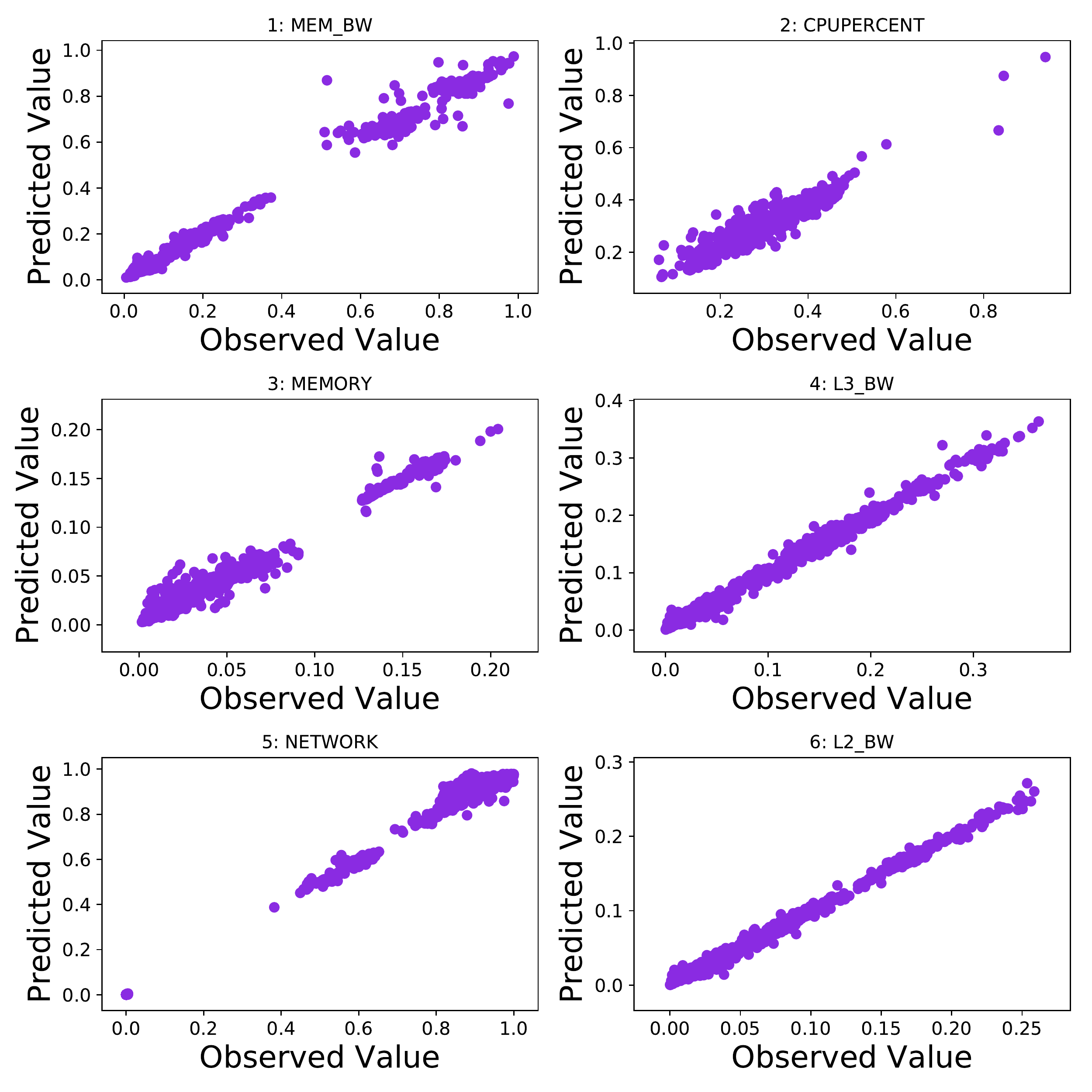}
\vspace*{-5mm}
\caption{Predicted resource utilizations across different resource
  types. Points concentrated along the diagonal indicate that the predicted
  values match the actual observed values.}
\label{fig:resource_prediction}
\end{figure}

Figure~\ref{fig:resource_prediction} shows the accuracy of FECBench in
predicting the actual resource utilizations for the co-located workloads.  We see that
the resource utilization predicted by FECBench are quite accurate as the bulk
of the points fall close to the diagonal region of the chart with very few outliers.

\subsection{Validating the Design Space Exploration Strategy}
\label{eval_designspace}

\begin{figure*}[t]
\centering
\includegraphics[width=\textwidth,height=4cm]{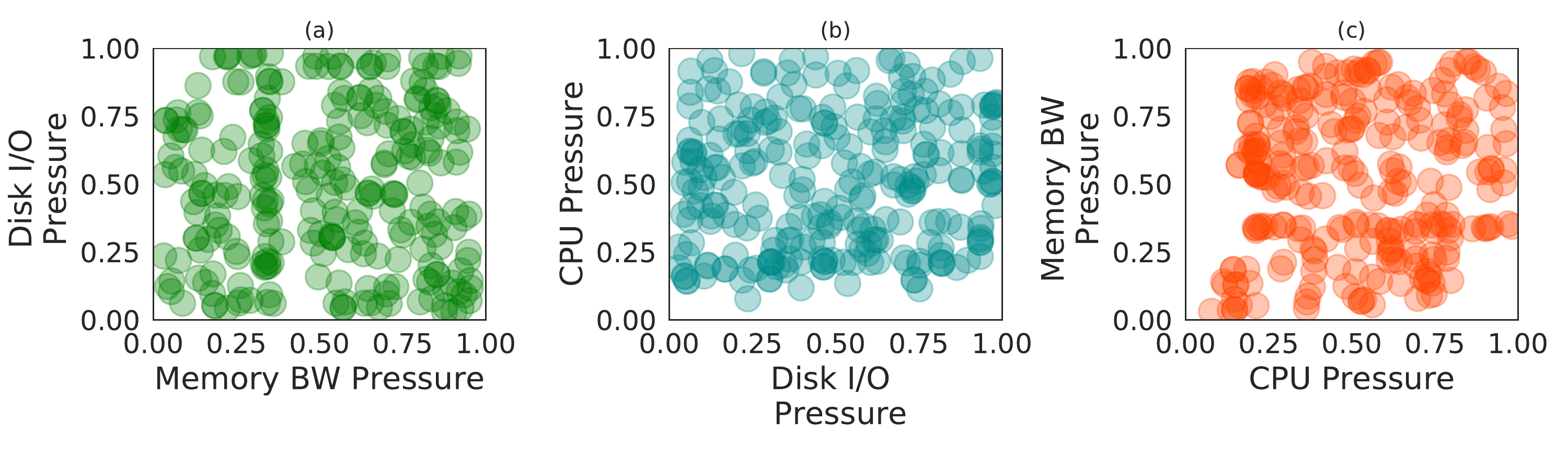}
\vspace*{-9mm}
\caption{Coverage of the resource stressor design space exerted by the co-located application combinations available in the knowledge base.}
\setlength{\textfloatsep}{0.7\baselineskip plus 0.2\baselineskip minus 0.5\baselineskip}
\label{fig:DOE_Experiment}
\end{figure*}

For this experiment, the goal is to find the right application combinations
that exert pressure in a  tunable fashion along multiple resource dimensions
(in our case, CPU, memory bandwidth and disk resources).  We set the number of
samples for the LHS strategy to 300 in the design of experiments, and got coverage for around 264
bins, i.e., 88\%. To allow for easier visualization of the coverage, we project the three dimensional datapoints on a two
dimensional scale as shown in Figure~\ref{fig:DOE_Experiment}, demonstrating
good coverage of the design space.

\subsection{Validating the Accuracy of the Performance Models}
\label{eval_perfmodel}

We used specific applications drawn from the DaCaPo benchmark, the Parsec benchmark
and the Keras machine learning application model as target applications whose performance models we were interested in.
Specifically, from
the DaCaPo benchmark, we chose PMD which is an application that analyzes large-scale Java source
code classes for source code problems.  From the Parsec benchmark, we chose the
Canneal application, which uses a cache-aware simulated annealing approach for
routing cost minimization in chip design.  The Canneal program has a very high
bandwidth requirement  and also large working
sets~\cite{bienia2008parsec}. From the Keras machine learning application
model, we used InceptionResnetV2, which represents an emerging workload class
for prediction inference serving systems~\cite{urlKeras}.

We first build performance models for these three applications using our
approach as discussed in Section~\ref{sec:methodology}. To test the
effectiveness of  the learned application performance models, we co-locate the
target applications with the web search workload from the CloudSuite
benchmark \cite{ferdman2012clearing}, which uses the Apache Solr search engine framework and
emulates varying number of clients that query this web search engine. We ran
three  different scenarios with varying number of clients: 600,  800, 900, and
1000.  We placed our target application with a co-located
web-search server on the host compute node. We assigned two cores to
the target application and the rest of the cores to the web-search server.

\begin{figure}[t]
\centering
\setlength{\textfloatsep}{0.7\baselineskip plus 0.2\baselineskip minus 0.5\baselineskip}
\includegraphics[width=\linewidth,height=4cm]{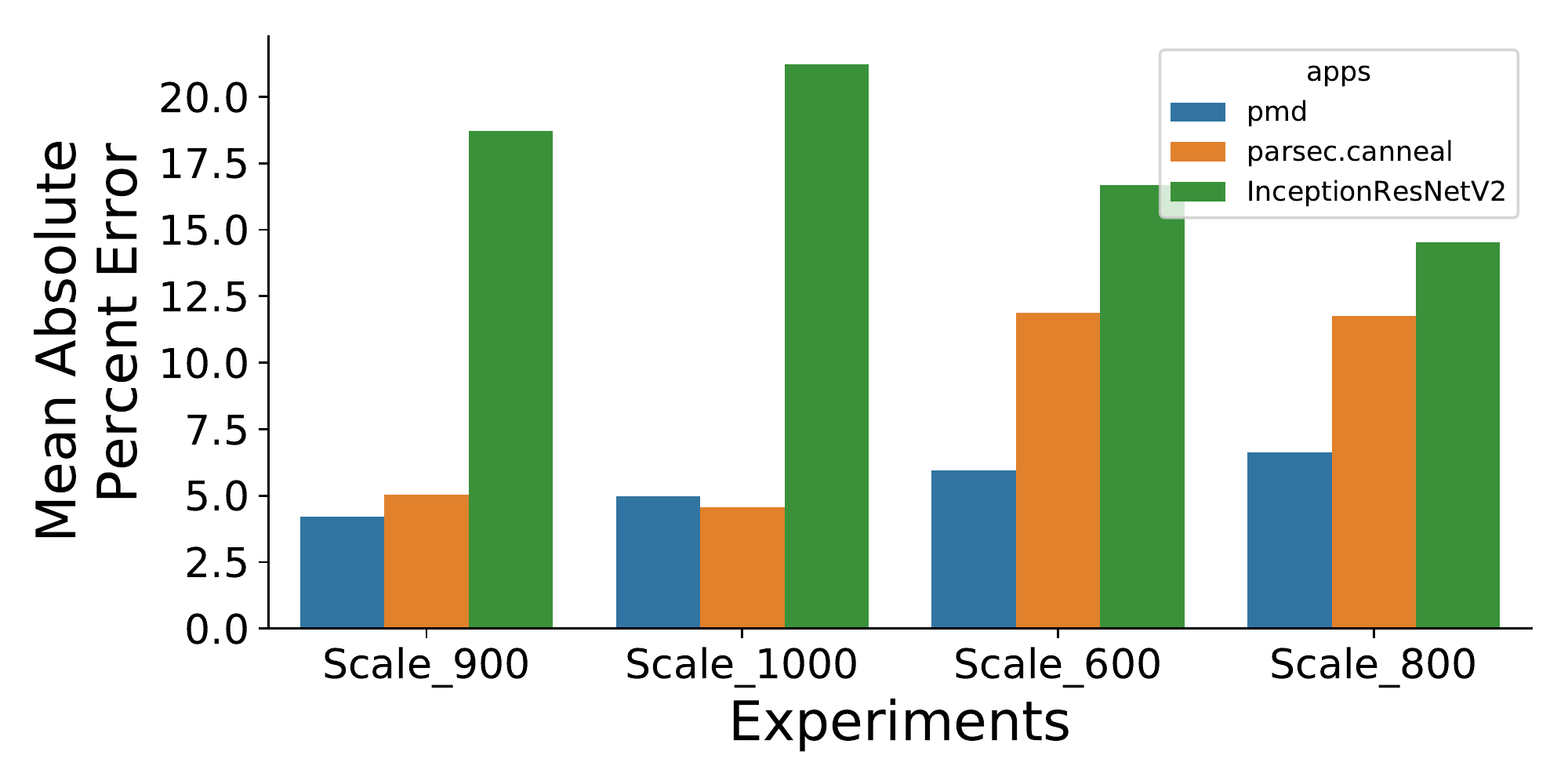}
\vspace*{-5mm}
\caption{Prediction accuracy in mean absolute percent error for \texttt{PMD},
\texttt{Canneal}, \texttt{InceptionResnetV2} applications when co-located
with web search server from CloudSuite.}
    \setlength{\belowcaptionskip}{-12pt}
\label{fig:eval_mape}
\end{figure}

\begin{figure}[t]
\centering
\setlength{\textfloatsep}{0.7\baselineskip plus 0.2\baselineskip minus 0.5\baselineskip}
\includegraphics[width=\linewidth]{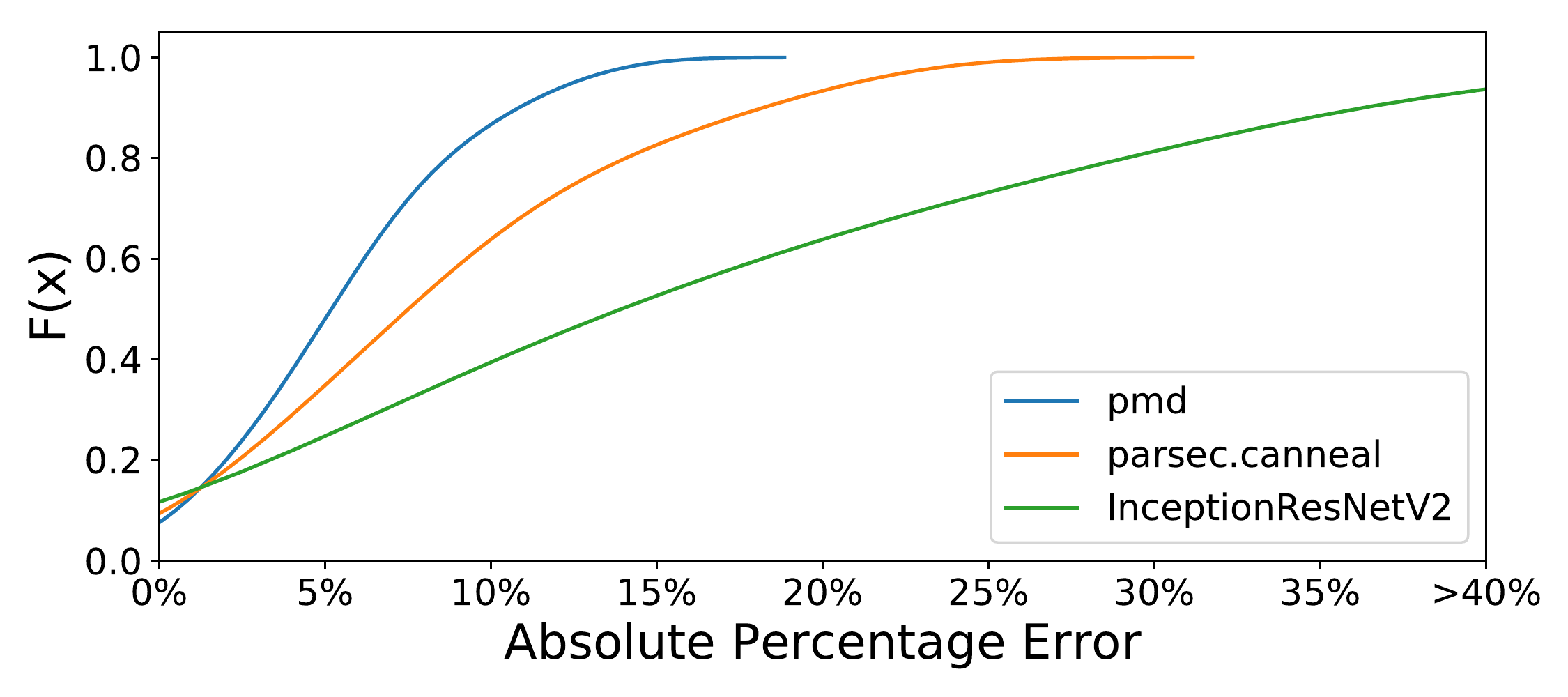}
\vspace*{-5mm}
\caption{Cumulative distributions of prediction errors for the \texttt{PMD},
\texttt{Canneal}, \texttt{InceptionResnetV2} applications.}
    \setlength{\belowcaptionskip}{-12pt}
\label{fig:eval_cdf}
\end{figure}

Figure~\ref{fig:eval_mape} shows the mean absolute percent errors (MAPEs) for the
three applications under varying degrees of loads generated by the
clients of the co-located web search application. For example, when the number
of clients is 800, the MAPEs are 6.6\%, 11.8\% and  14.5\%
for PMD, Canneal and InceptionResnetV2, respectively. Also, the median percentage errors
for all the cases are below 5.4\%,  7.6\% and 13.5\% for PMD, Canneal and
InceptionResnetV2, respectively.  To showcase the total number of correct
predictions made by the system, we leverage a CDF curve that has been
used in the literature to showcase the effectiveness of the machine learning
models \cite{morris2018model}. Figure \ref{fig:eval_cdf} illustrates that, for
the PMD application, 80\% of the predictions have error rates less than
9\%. For the Canneal application, 80\% of the predictions have error rates
below 15\%. For the InceptionResNetV2 application, about 70\% of the predictions have error rates below 25\%.

\subsection{FECBench in Action: A Concrete Use Case}
\label{sec:validate}

Besides validating the efficacy of FECBench on applications drawn from the 
benchmarking suites, we have applied FECBench to interference-aware
load balancing of topics for a publish-process-subscribe system~\cite{PubSub_SEC:18}.
The Publish/Subscribe (pub/sub) communication pattern allows asynchronous and
anonymous exchange of information (topic of interest) between publishers
(data producers) and subscribers (data receivers). Therefore, pub/sub is
widely used to meet the scalable data distribution needs of IoT applications,
where large amounts of data produced by sensors are distributed and
processed by receivers for closed-loop actuation. The need for
processing sensor data is accomplished on broker nodes that
route information between publishers and subscribers.

\begin{figure}[t]
    \centering
    \vspace*{-5mm}
    \subfloat[]{{\includegraphics[width=0.45\linewidth,height=1.5in]{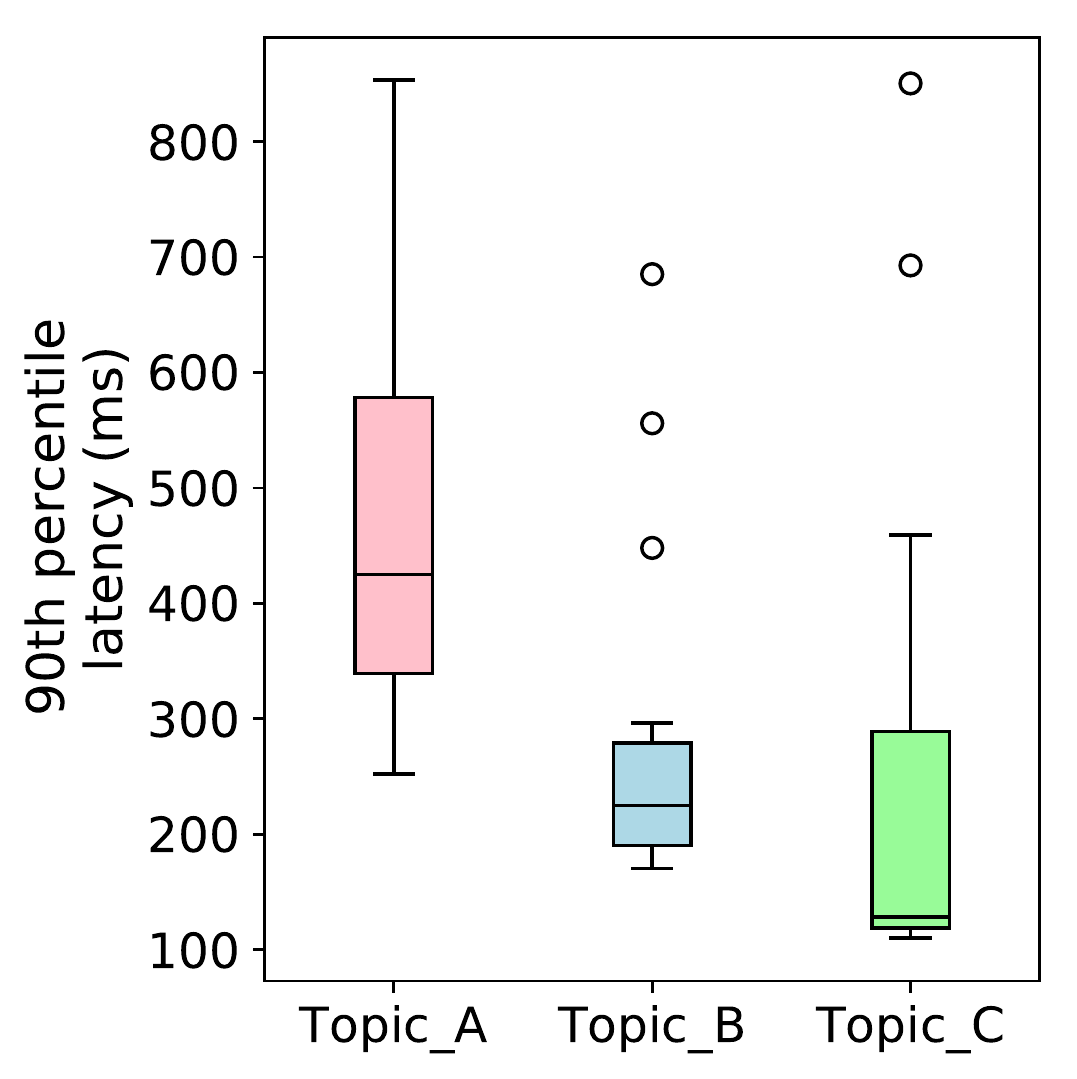} }}%
    \subfloat[]{{\includegraphics[width=0.45\linewidth,height=1.5in]{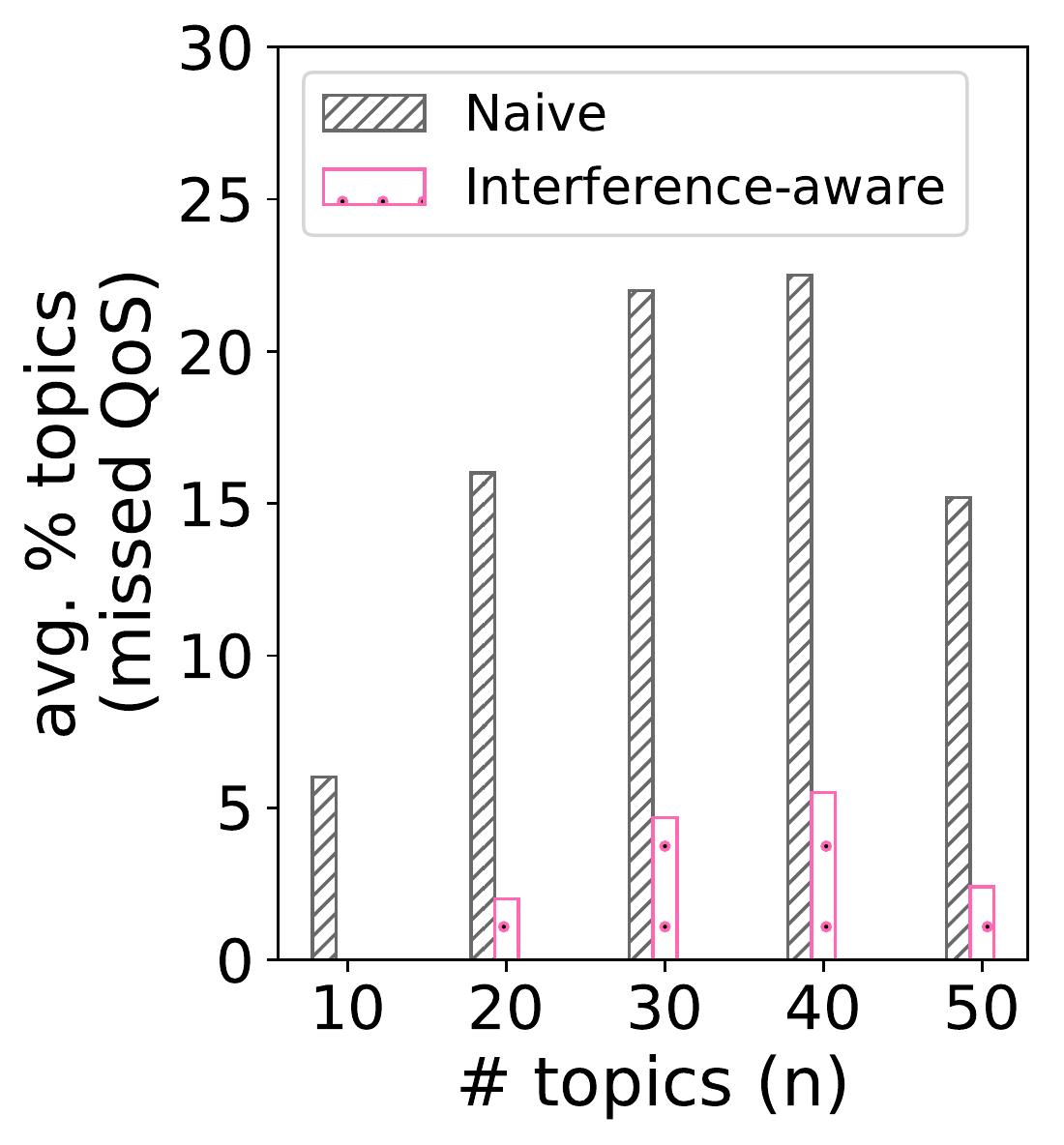} }}%
    \caption{Use of FECBench in publish-process-subscribe showing: (a) impact of
      co-location on a topic's latency; (b) interference-aware placement of
      topics.}%
    \label{fig:pubsub}%
\end{figure}

In a publish-process-subscribe system, a topic's latency can suffer
significantly due to the processing demands of other co-located topics at the same
broker. Figure~\ref{fig:pubsub} demonstrates this effect. 
Here, a topic is characterized by its processing interval $p$, i.e.,
average time for processing each incoming message on the topic, and cumulative
publishing rate $r$, at which messages arrive at the
topic. Figure~\ref{fig:pubsub}(a) shows that topics A, B and C
show wide variations in their 90th percentile latencies
($\sim100$ms to $\sim800$ms) under varying background loads.

%

For latency critical IoT applications, it is necessary to ensure that a topic's
latency is within a desirable QoS value. Therefore, it is important to
co-locate topics at the brokers in an interference-aware manner such that none
of the topics in the system violate their latency QoS. To this end, one
approach is to learn a latency prediction model for the brokers in the
pub/sub system by leveraging the FECBench approach.
Subsequently, the latency prediction model can be used to determine
which topics can be safely co-located at a broker without incurring QoS
violations. Figure~\ref{fig:pubsub}(b) shows how such an interference-aware
method can reduce the percentage of topics in the system that suffer from QoS
violations. Here, the interference-aware approach,
which uses the latency prediction model
obtained by the FECBench approach, is able to meet the QoS for $\sim95\%$ of the topics in the
system. This is significantly better than a naive approach based on round robin scheduling,
which is only able to meet the QoS for $\sim80\%$ of the topics in the system.


\section{Conclusion}
\label{sec:conclusions}

Making effective dynamic resource management decisions to maintain application
service level objectives (SLOs) in multi-tenant cloud platforms including the
emerging fog/edge environments requires an accurate understanding of the 
application performance in the presence of different levels of performance
interference that an application is likely to encounter when co-located with other
workloads.  Data-driven performance models can capture
these properties, which in turn can be used in a feedback loop to make
effective resource management decisions.  The vast number of applications,
their co-location patterns, differences in their workload types, platform
heterogeneity and an overall lack of a systematic performance model building
framework make it an extremely daunting task for developers to build such
performance models. FECBench (Fog/Edge/Cloud Benchmarking) is a framework that
addresses these challenges, thereby relieving the developers from expending
significant time and effort, and incurring prohibitive costs in this process. 
It provides an extensible resource monitoring and metrics collection
capability, a collection of disparate benchmarks integrated within a single
framework, and a systematic and scalable model building process with
an extensible knowledge base application combinations that create resource stress across the multi-dimensional
resources design space. Empirical evaluations on different
application use cases demonstrate that the predicted application performance using
the FECBench approach incurs a median error of only 7.6\%
across all test cases, with 5.4\% in the best case and 13.5\% in the
worst case.  FECBench is available in open source at \url{github.com/doc-vu}.

So far, FECBench has been evaluated on a single hardware platform. Its efficacy needs
to be validated on a variety of hardware platforms. To that end, we will
explore the use of transfer learning to minimize the efforts.  Our use of 106
applications did not provide coverage across every possible resource dimension
and hence improving the coverage is another area of future work.


\section*{Acknowledgments}

This work is supported in part by NSF US Ignite CNS 1531079 and AFOSR
DDDAS FA9550-18-1-0126 and NIST 70NANB17H274 and  AFRL/Lockheed Martin's StreamlinedML program.  Any opinions, findings, and conclusions or
recommendations expressed in this material are those of the author(s) and do
not necessarily reflect the views of these funding agencies.



\bibliography{fecbench}

\begin{thebibliography}{10}
\providecommand{\url}[1]{#1}
\csname url@samestyle\endcsname
\providecommand{\newblock}{\relax}
\providecommand{\bibinfo}[2]{#2}
\providecommand{\BIBentrySTDinterwordspacing}{\spaceskip=0pt\relax}
\providecommand{\BIBentryALTinterwordstretchfactor}{4}
\providecommand{\BIBentryALTinterwordspacing}{\spaceskip=\fontdimen2\font plus
\BIBentryALTinterwordstretchfactor\fontdimen3\font minus
  \fontdimen4\font\relax}
\providecommand{\BIBforeignlanguage}[2]{{%
\expandafter\ifx\csname l@#1\endcsname\relax
\typeout{** WARNING: IEEEtran.bst: No hyphenation pattern has been}%
\typeout{** loaded for the language `#1'. Using the pattern for}%
\typeout{** the default language instead.}%
\else
\language=\csname l@#1\endcsname
\fi
#2}}
\providecommand{\BIBdecl}{\relax}
\BIBdecl

\bibitem{xavier2016understanding}
M.~G. Xavier, K.~J. Matteussi, F.~Lorenzo, and C.~A. De~Rose, ``Understanding
  performance interference in multi-tenant cloud databases and web
  applications,'' in \emph{Big Data (Big Data), 2016 IEEE International
  Conference on}.\hskip 1em plus 0.5em minus 0.4em\relax IEEE, 2016, pp.
  2847--2852.

\bibitem{iosup2011performance}
A.~Iosup, N.~Yigitbasi, and D.~Epema, ``On the performance variability of
  production cloud services,'' in \emph{Cluster, Cloud and Grid Computing
  (CCGrid), 2011 11th IEEE/ACM International Symposium on}.\hskip 1em plus
  0.5em minus 0.4em\relax IEEE, 2011, pp. 104--113.

\bibitem{delimitrou2013paragon}
C.~Delimitrou and C.~Kozyrakis, ``Paragon: Qos-aware scheduling for
  heterogeneous datacenters,'' in \emph{ACM SIGPLAN Notices}, vol.~48,
  no.~4.\hskip 1em plus 0.5em minus 0.4em\relax ACM, 2013, pp. 77--88.

\bibitem{berekmeri2016feedback}
M.~Berekmeri, D.~Serrano, S.~Bouchenak, N.~Marchand, and B.~Robu, ``Feedback
  autonomic provisioning for guaranteeing performance in mapreduce systems,''
  \emph{IEEE Transactions on Cloud Computing}, 2016.

\bibitem{rattihalli2018exploring}
G.~Rattihalli, ``Exploring potential for resource request right-sizing via
  estimation and container migration in apache mesos,'' in \emph{2018 IEEE/ACM
  International Conference on Utility and Cloud Computing Companion (UCC
  Companion)}.\hskip 1em plus 0.5em minus 0.4em\relax IEEE, 2018, pp. 59--64.

\bibitem{delvalle2016exploring}
R.~DelValle, G.~Rattihalli, A.~Beltre, M.~Govindaraju, and M.~J. Lewis,
  ``Exploring the design space for optimizations with apache aurora and
  mesos,'' in \emph{2016 IEEE 9th International Conference on Cloud Computing
  (CLOUD)}.\hskip 1em plus 0.5em minus 0.4em\relax IEEE, 2016, pp. 537--544.

\bibitem{xu2018pythia}
R.~Xu, S.~Mitra, J.~Rahman, P.~Bai, B.~Zhou, G.~Bronevetsky, and S.~Bagchi,
  ``Pythia: Improving datacenter utilization via precise contention prediction
  for multiple co-located workloads,'' in \emph{Proceedings of the 19th
  International Middleware Conference}.\hskip 1em plus 0.5em minus 0.4em\relax
  ACM, 2018, pp. 146--160.

\bibitem{novakovic2013deepdive}
D.~Novakovic, N.~Vasic, S.~Novakovic, D.~Kostic, and R.~Bianchini, ``Deepdive:
  Transparently identifying and managing performance interference in
  virtualized environments,'' in \emph{Proceedings of the 2013 USENIX Annual
  Technical Conference}, no. EPFL-CONF-185984, 2013.

\bibitem{urlstressng}
\BIBentryALTinterwordspacing
(1999) Stress-ng. [Online]. Available:
  \url{http://kernel.ubuntu.com/~cking/stress-ng/}
\BIBentrySTDinterwordspacing

\bibitem{brogi2017qos}
A.~Brogi and S.~Forti, ``Qos-aware deployment of iot applications through the
  fog,'' \emph{IEEE Internet of Things Journal}, vol.~4, no.~5, pp. 1185--1192,
  2017.

\bibitem{jonathan2017nebula}
A.~Jonathan, M.~Ryden, K.~Oh, A.~Chandra, and J.~Weissman, ``Nebula:
  Distributed edge cloud for data intensive computing,'' \emph{IEEE
  Transactions on Parallel and Distributed Systems}, vol.~28, no.~11, pp.
  3229--3242, 2017.

\bibitem{iosup2014iaas}
A.~Iosup, R.~Prodan, and D.~Epema, ``Iaas cloud benchmarking: approaches,
  challenges, and experience,'' in \emph{Cloud Computing for Data-Intensive
  Applications}.\hskip 1em plus 0.5em minus 0.4em\relax Springer, 2014, pp.
  83--104.

\bibitem{oleksenko2017fex}
O.~Oleksenko, D.~Kuvaiskii, P.~Bhatotia, and C.~Fetzer, ``Fex: A software
  systems evaluator,'' in \emph{Dependable Systems and Networks (DSN), 2017
  47th Annual IEEE/IFIP International Conference on}.\hskip 1em plus 0.5em
  minus 0.4em\relax IEEE, 2017, pp. 543--550.

\bibitem{barve2018fecbench}
Y.~Barve, S.~Shekhar, A.~Chhokra, S.~Khare, A.~Bhattacharjee, and A.~Gokhale,
  ``Fecbench: An extensible framework for pinpointing sources of performance
  interference in the cloud-edge resource spectrum,'' in \emph{2018 IEEE/ACM
  Symposium on Edge Computing (SEC)}.\hskip 1em plus 0.5em minus 0.4em\relax
  IEEE, 2018, pp. 331--333.

\bibitem{urlKeras}
\BIBentryALTinterwordspacing
(2018) Keras - inceptionresnetv2. [Online]. Available:
  \url{https://keras.io/applications/#inceptionresnetv2}
\BIBentrySTDinterwordspacing

\bibitem{mars2011bubble}
J.~Mars, L.~Tang, R.~Hundt, K.~Skadron, and M.~L. Soffa, ``Bubble-up:
  Increasing utilization in modern warehouse scale computers via sensible
  co-locations,'' in \emph{Proceedings of the 44th annual IEEE/ACM
  International Symposium on Microarchitecture}.\hskip 1em plus 0.5em minus
  0.4em\relax ACM, 2011, pp. 248--259.

\bibitem{yang2013bubble}
H.~Yang, A.~Breslow, J.~Mars, and L.~Tang, ``Bubble-flux: Precise online qos
  management for increased utilization in warehouse scale computers,'' in
  \emph{ACM SIGARCH Computer Architecture News}, vol.~41, no.~3.\hskip 1em plus
  0.5em minus 0.4em\relax ACM, 2013, pp. 607--618.

\bibitem{zhao2013empirical}
J.~Zhao, H.~Cui, J.~Xue, X.~Feng, Y.~Yan, and W.~Yang, ``An empirical model for
  predicting cross-core performance interference on multicore processors,'' in
  \emph{Proceedings of the 22nd international conference on Parallel
  architectures and compilation techniques}.\hskip 1em plus 0.5em minus
  0.4em\relax IEEE Press, 2013, pp. 201--212.

\bibitem{javadi2017dial}
S.~A. Javadi and A.~Gandhi, ``Dial: Reducing tail latencies for cloud
  applications via dynamic interference-aware load balancing,'' in
  \emph{Autonomic Computing (ICAC), 2017 IEEE International Conference
  on}.\hskip 1em plus 0.5em minus 0.4em\relax IEEE, 2017, pp. 135--144.

\bibitem{subramanian2015application}
L.~Subramanian, V.~Seshadri, A.~Ghosh, S.~Khan, and O.~Mutlu, ``The application
  slowdown model: Quantifying and controlling the impact of inter-application
  interference at shared caches and main memory,'' in \emph{Proceedings of the
  48th International Symposium on Microarchitecture}.\hskip 1em plus 0.5em
  minus 0.4em\relax ACM, 2015, pp. 62--75.

\bibitem{mishra2017esp}
N.~Mishra, J.~D. Lafferty, and H.~Hoffmann, ``Esp: A machine learning approach
  to predicting application interference,'' in \emph{Autonomic Computing
  (ICAC), 2017 IEEE International Conference on}.\hskip 1em plus 0.5em minus
  0.4em\relax IEEE, 2017, pp. 125--134.

\bibitem{govindan2011cuanta}
S.~Govindan, J.~Liu, A.~Kansal, and A.~Sivasubramaniam, ``Cuanta: quantifying
  effects of shared on-chip resource interference for consolidated virtual
  machines,'' in \emph{Proceedings of the 2nd ACM Symposium on Cloud
  Computing}.\hskip 1em plus 0.5em minus 0.4em\relax ACM, 2011, p.~22.

\bibitem{delimitrou2013ibench}
C.~Delimitrou and C.~Kozyrakis, ``ibench: Quantifying interference for
  datacenter applications,'' in \emph{2013 IEEE international symposium on
  workload characterization (IISWC)}.\hskip 1em plus 0.5em minus 0.4em\relax
  IEEE, 2013, pp. 23--33.

\bibitem{pu2013your}
X.~Pu, L.~Liu, Y.~Mei, S.~Sivathanu, Y.~Koh, C.~Pu, and Y.~Cao, ``Who is your
  neighbor: Net i/o performance interference in virtualized clouds,''
  \emph{IEEE Transactions on Services Computing}, vol.~6, no.~3, pp. 314--329,
  2013.

\bibitem{silva2013cloudbench}
M.~Silva, M.~R. Hines, D.~Gallo, Q.~Liu, K.~D. Ryu, and D.~Da~Silva,
  ``Cloudbench: Experiment automation for cloud environments,'' in \emph{Cloud
  Engineering (IC2E), 2013 IEEE International Conference on}.\hskip 1em plus
  0.5em minus 0.4em\relax IEEE, 2013, pp. 302--311.

\bibitem{jayathilaka2018detecting}
H.~Jayathilaka, C.~Krintz, and R.~M. Wolski, ``Detecting performance anomalies
  in cloud platform applications,'' \emph{IEEE Transactions on Cloud
  Computing}, no.~1, pp. 1--1, 2018.

\bibitem{urlCollectd}
\BIBentryALTinterwordspacing
(2018) Collectd - the system statistics collection daemon. [Online]. Available:
  \url{https://collectd.org/}
\BIBentrySTDinterwordspacing

\bibitem{anirbanUCC}
A.~{Bhattacharjee}, Y.~{Barve}, A.~{Gokhale}, and T.~{Kuroda}, ``A model-driven
  approach to automate the deployment and management of cloud services,'' in
  \emph{2018 IEEE/ACM International Conference on Utility and Cloud Computing
  Companion (UCC Companion)}, Dec 2018, pp. 109--114.

\bibitem{barve2018pads}
Y.~D. Barve, P.~Patil, A.~Bhattacharjee, and A.~Gokhale, ``Pads: Design and
  implementation of a cloud-based, immersive learning environment for
  distributed systems algorithms,'' \emph{IEEE Transactions on Emerging Topics
  in Computing}, vol.~6, no.~1, pp. 20--31, 2018.

\bibitem{barve2018upsara}
Y.~Barve, S.~Shekhar, S.~Khare, A.~Bhattacharjee, and A.~Gokhale, ``Upsara: A
  model-driven approach for performance analysis of cloud-hosted
  applications,'' in \emph{2018 IEEE/ACM 11th International Conference on
  Utility and Cloud Computing (UCC)}.\hskip 1em plus 0.5em minus 0.4em\relax
  IEEE, 2018, pp. 1--10.

\bibitem{bienia2008parsec}
C.~Bienia, S.~Kumar, J.~P. Singh, and K.~Li, ``The parsec benchmark suite:
  Characterization and architectural implications,'' in \emph{Proceedings of
  the 17th international conference on Parallel architectures and compilation
  techniques}.\hskip 1em plus 0.5em minus 0.4em\relax ACM, 2008, pp. 72--81.

\bibitem{blackburn2006dacapo}
S.~M. Blackburn, R.~Garner, C.~Hoffmann, A.~M. Khang, K.~S. McKinley,
  R.~Bentzur, A.~Diwan, D.~Feinberg, D.~Frampton, S.~Z. Guyer \emph{et~al.},
  ``The dacapo benchmarks: Java benchmarking development and analysis,'' in
  \emph{ACM Sigplan Notices}, vol.~41, no.~10.\hskip 1em plus 0.5em minus
  0.4em\relax ACM, 2006, pp. 169--190.

\bibitem{urlPhoronix}
\BIBentryALTinterwordspacing
(2018) Phoronix benchmark. [Online]. Available:
  \url{http://www.phoronix-test-suite.com/}
\BIBentrySTDinterwordspacing

\bibitem{yazdanbakhsh2017axbench}
A.~Yazdanbakhsh, D.~Mahajan, H.~Esmaeilzadeh, and P.~Lotfi-Kamran, ``Axbench: A
  multiplatform benchmark suite for approximate computing,'' \emph{IEEE Design
  \& Test}, vol.~34, no.~2, pp. 60--68, 2017.

\bibitem{mishra2010towards}
A.~K. Mishra, J.~L. Hellerstein, W.~Cirne, and C.~R. Das, ``Towards
  characterizing cloud backend workloads: insights from google compute
  clusters,'' \emph{ACM SIGMETRICS Performance Evaluation Review}, vol.~37,
  no.~4, pp. 34--41, 2010.

\bibitem{rousseeuw1987silhouettes}
P.~J. Rousseeuw, ``Silhouettes: a graphical aid to the interpretation and
  validation of cluster analysis,'' \emph{Journal of computational and applied
  mathematics}, vol.~20, pp. 53--65, 1987.

\bibitem{cavazzuti2013design}
M.~Cavazzuti, ``Design of experiments,'' in \emph{Optimization Methods}.\hskip
  1em plus 0.5em minus 0.4em\relax Springer, 2013, pp. 13--42.

\bibitem{loh1996latin}
W.-L. Loh \emph{et~al.}, ``On latin hypercube sampling,'' \emph{The annals of
  statistics}, vol.~24, no.~5, pp. 2058--2080, 1996.

\bibitem{quinlan1986induction}
J.~R. Quinlan, ``Induction of decision trees,'' \emph{Machine learning},
  vol.~1, no.~1, pp. 81--106, 1986.

\bibitem{barve2016cloud}
Y.~D. Barve, P.~Patil, and A.~Gokhale, ``A cloud-based immersive learning
  environment for distributed systems algorithms,'' in \emph{2016 IEEE 40th
  Annual Computer Software and Applications Conference (COMPSAC)},
  vol.~1.\hskip 1em plus 0.5em minus 0.4em\relax IEEE, 2016, pp. 754--763.

\bibitem{urlLikwid}
\BIBentryALTinterwordspacing
(2018) Likwid - system monitoring tool. [Online]. Available:
  \url{https://github.com/RRZE-HPC/likwid}
\BIBentrySTDinterwordspacing

\bibitem{urlInfluxdb}
\BIBentryALTinterwordspacing
(2018) Influxdb - time series database. [Online]. Available:
  \url{https://www.influxdata.com/time-series-platform/influxdb/}
\BIBentrySTDinterwordspacing

\bibitem{urlScikit}
\BIBentryALTinterwordspacing
(2018) Scikit-learn - machine learning library in python. [Online]. Available:
  \url{https://scikit-learn.org/stable/}
\BIBentrySTDinterwordspacing

\bibitem{ferdman2012clearing}
M.~Ferdman, A.~Adileh, O.~Kocberber, S.~Volos, M.~Alisafaee, D.~Jevdjic,
  C.~Kaynak, A.~D. Popescu, A.~Ailamaki, and B.~Falsafi, ``Clearing the clouds:
  a study of emerging scale-out workloads on modern hardware,'' in \emph{ACM
  SIGPLAN Notices}, vol.~47, no.~4.\hskip 1em plus 0.5em minus 0.4em\relax ACM,
  2012, pp. 37--48.

\bibitem{morris2018model}
N.~Morris, C.~Stewart, L.~Chen, R.~Birke, and J.~Kelley, ``Model-driven
  computational sprinting,'' in \emph{Proceedings of the Thirteenth EuroSys
  Conference}.\hskip 1em plus 0.5em minus 0.4em\relax ACM, 2018, p.~38.

\bibitem{PubSub_SEC:18}
S.~Khare, H.~Sun, K.~Zhang, J.~Gascon-Samson, A.~Gokhale, X.~Koutsoukos, and
  H.~Abdelaziz, ``Scalable edge computing for low latency data dissemination in
  topic-based publish/subscribe,'' in \emph{2018 IEEE/ACM Symposium on Edge
  Computing (SEC)}.\hskip 1em plus 0.5em minus 0.4em\relax IEEE, 2018, pp.
  214--227.

\end{thebibliography}
\bibliographystyle{IEEEtran}

\end{document}